\documentclass[aps,prx,twocolumn,hyperlinks,superscriptaddress,floatfix,showpacs,citeautoscript,longbibliography,hyperlinks]{revtex4-1}
\usepackage{amsmath}
\usepackage{amssymb}
\usepackage{amsfonts}
\usepackage{dsfont}
\usepackage{color}
\usepackage{nicefrac}
\usepackage[autostyle]{csquotes}
\usepackage[colorlinks=true,citecolor=blue]{hyperref}
\usepackage{graphicx,graphics}
\usepackage{siunitx}
\usepackage{braket}
\usepackage{float}
\usepackage[dvipsnames]{xcolor}
\usepackage{xcolor}
\usepackage{comment}
\usepackage[T1]{fontenc}

\usepackage{soul}

\begin{document}

\title{Dephasing in a crystal-phase defined double quantum dot charge qubit strongly coupled to a high-impedance resonator}

\author{Antti Ranni}
\email{antti.ranni@ftf.lth.se}
\author{Subhomoy Haldar}
\author{Harald Havir}
\author{Sebastian Lehmann}
\affiliation{NanoLund and Solid State Physics, Lund University, Box 118, 22100 Lund, Sweden}
\author{Pasquale Scarlino}
\affiliation{Institute of Physics and Center for Quantum Science and Engineering,
Ecole Polytechnique F\'{e}d\'{e}rale de Lausanne, CH-1015 Lausanne, Switzerland}
\author{Andreas Baumgartner}
\author{Christian Schönenberger}
\affiliation{Department of Physics and Swiss Nanoscience Institute, University of Basel, Klingelbergstrasse 82 CH-4056, Switzerland}
%\affiliation{Swiss Nanoscience Institute, University of Basel, Klingelbergstrasse 82 CH-4056, Switzerland}
\author{Claes Thelander}
\affiliation{NanoLund and Solid State Physics, Lund University, Box 118, 22100 Lund, Sweden}
\author{Kimberly A. Dick}
\affiliation{NanoLund and Solid State Physics, Lund University, Box 118, 22100 Lund, Sweden}
\affiliation{Center for Analysis and Synthesis, Lund University, Box 124, 22100 Lund, Sweden}
\author{Patrick P. Potts}
\affiliation{Department of Physics and Swiss Nanoscience Institute, University of Basel, Klingelbergstrasse 82 CH-4056, Switzerland}
\author{Ville F. Maisi}
\email{ville.maisi@ftf.lth.se}
\affiliation{NanoLund and Solid State Physics, Lund University, Box 118, 22100 Lund, Sweden}

\date{\today}

\begin{abstract}
    Dephasing of a charge qubit is usually credited to charge noise in the environment. Here we show that charge noise may not be the limiting factor for the qubit coherence. To this end, we study coherence properties of a crystal-phase defined semiconductor nanowire double quantum dot (DQD) charge qubit strongly coupled to a high-impedance resonator using radio-frequency (RF) reflectometry. Response of this hybrid system is measured both at a charge noise sensitive operation point (with finite DQD detuning) and at an insensitive point (so-called sweet spot with zero detuning). A theoretical model based on Jaynes-Cummings Hamiltonian matches the experimental results well and yields only a 10 \% difference in dephasing rates between the two cases, despite that the sensitivity to detuning charge noise differs by a factor of 5. Therefore the charge noise is not the limiting factor for the coherence in this type of semiconducting nanowire qubits.  
\end{abstract}

\maketitle
%\linenumbers

\section{INTRODUCTION}

The interplay between photons stored in a cavity and a coherent quantum system, here a qubit, in the framework of cavity quantum electrodynamics (QED) can give rise to a coherent light-matter interaction and serve as a platform for circuit QED experiments~\cite{Haroche2006}. The strength of this interaction is characterized by the coupling rate $g$. If $g$ is larger than the cavity losses $\kappa$ and the total dephasing rate $\Gamma$ of the qubit, the photonic and electronic states are said to be strongly coupled. Strong coupling has been demonstrated in various quantum optics systems such as alkali atoms~\cite{Thompson1992}, Rydberg atoms~\cite{Brune1996}, superconducting qubits~\cite{Wallraff2004} and optically probed semiconductor single quantum dots in pholuminescence studies~\cite{Reithmaier2004,Yoshie2004}. These systems have attracted interest in the field of quantum information technology as they can be utilized for example, to coherently couple remote qubits~\cite{Majer2007,Sillanpaa2007} and for transferring quantum information from qubits to photons~\cite{Houck2007,Eichler2012}. In recent years the list of systems reaching the strong coupling limit has been extended to include semiconductor DQDs addressed with microwaves~\cite{Stockklauser2017,Mi2017,Bruhat2018,Scarlino2022,Yu2023}. In these mesoscopic solid-state devices microwave photons in a superconducting cavity interact with the electric dipole moment of a DQD charge qubit. While earlier studies with semiconductor quantum dots have demonstrated the strong coupling limit, the study of the dephasing of a charge qubit in this important limit has received less attention. In this work, we show that in the studied nanowire charge qubit with in-situ grown quantum dots, charge noise is not the dominant dephasing mechanism in the strong coupling limit. We reach the strong coupling between a crystal-phase defined DQD charge qubit in an InAs polytype nanowire~\cite{Nilsson2016,Barker2019} and a high-impedance Josephson junction array resonator~\cite{Masluk2012,Ranni2023}. In this limit, we investigate the charge qubit dephasing by measuring the RF response of the hybrid device at different resonant points where the cavity resonance and the gate-tunable qubit frequencies either match at finite DQD detuning or at sweet spot with zero detuning and a reduced sensitivity to external electric fields. The experimental data demonstrate comparable dephasing rates at the two operation points, in spite of a five times smaller charge noise sensitivity at the sweet spot than at finite detuning in excellent agreement with a theoretical model based on Jaynes-Cummings Hamiltonian. These experiments thus exclude charge noise as the main source of dephasing for our nanowire qubit.

\begin{figure*}
    \centering
	\includegraphics[width=\textwidth]{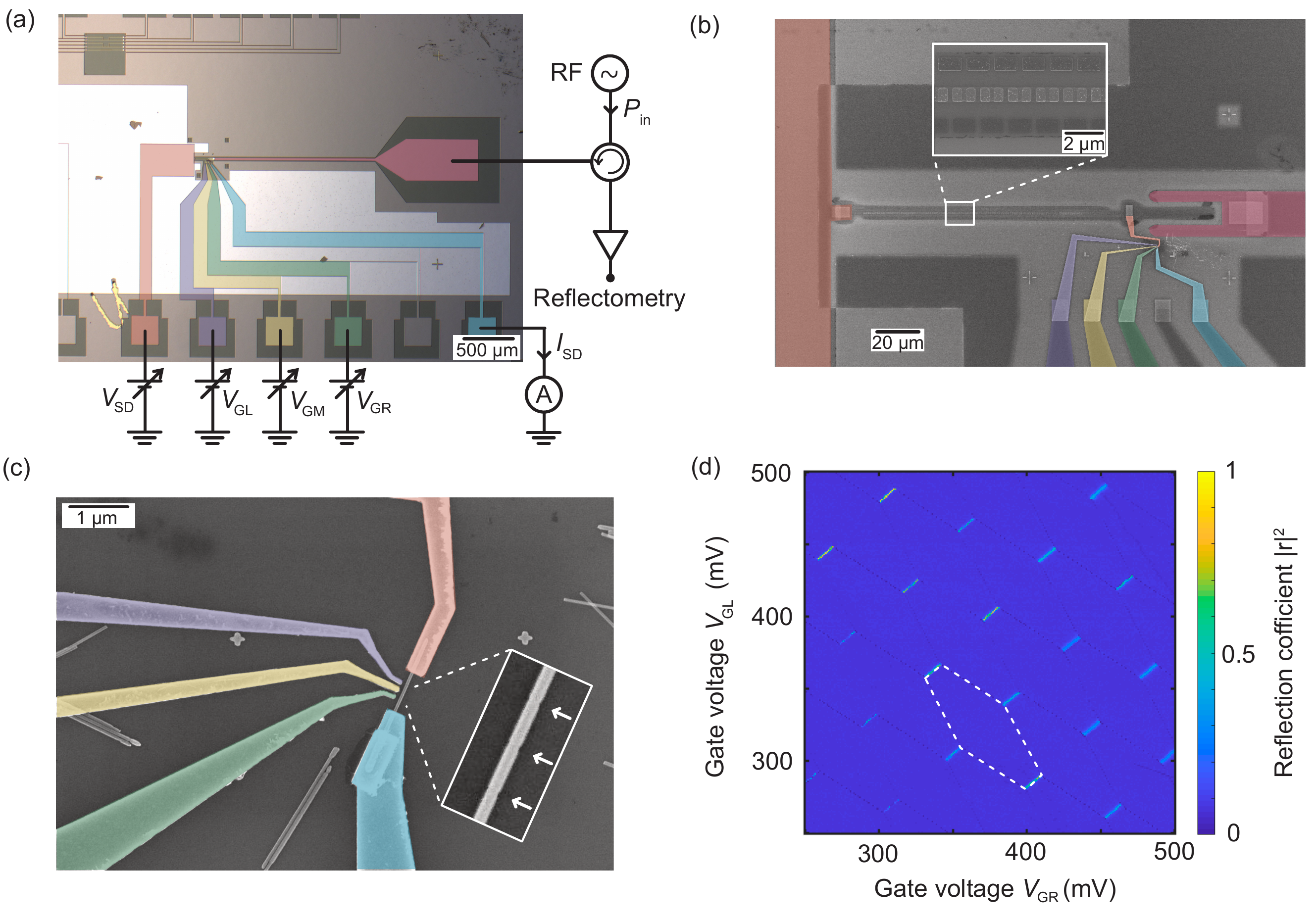}
	\caption{(a), An optical image highlighting the DC lines and the RF line of our resonator-DQD hybrid device. The light gray aluminum ground plane on top of the DC lines increases the line capacitance to reduce photon losses. Voltage $V_{\mathrm{SD}}$ biases the DQD. $V_{\mathrm{GL}}$, $V_{\mathrm{GR}}$ and $V_{\mathrm{GM}}$ move the chemical potentials of the quantum dots and tune the transparency of the tunnel barriers. The current $I_{\mathrm{SD}}$ is measured at the drain. (b), A scanning electron micrograph of the Josephson junction array resonator. The left side (voltage node) of the resonator is galvanically connected to the DC voltage bias line (large orange contact). The right side of the resonator is capacitively coupled to microwave feedline (in red) to probe our system by RF reflectometry. This voltage anti-node point is also galvanically connected to the DQD source contact (by the thin orange line). (c), A zoom-in showing the nanowire with the gate electrodes and source-drain contacts. The middle gate is aligned to point to the interdot barrier and the two other gates the middle points of the quantum dots. The inset shows a zoom-in of the nanowire before contact fabrication and the white arrows indicate the three barriers defining the DQD. (d), Reflection coefficient $|r|^{2}$ measured as a function of the left and the right gate voltages. The response is a honeycomb pattern for DQD where the bright lines indicate interdot transition lines. One of the DQD charge states is enclosed by the white dashed hexagon.
	\label{fig1} 
		}
\end{figure*}

\begin{figure*}
    \centering
	\includegraphics[width=\textwidth]{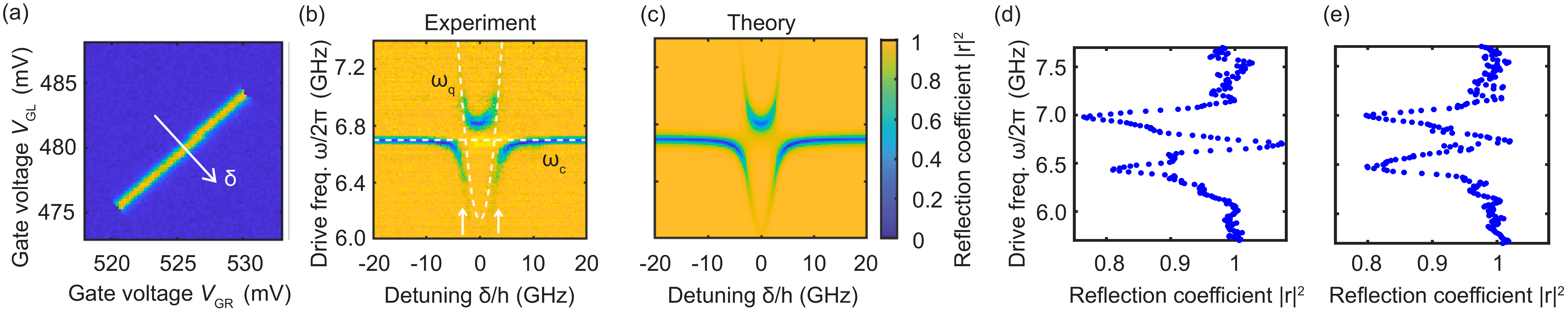}
	\caption{(a), The measured interdot line at fixed $V_{\mathrm{GM}}=\SI{150}{mV}$. The white line shows the detuning $\delta$ axis. (b), Measured $|r|^{2}$ as a function of the drive frequency $\omega$ and the detuning along the line shown in panel (a). The dashed white lines indicate $\omega_{\mathrm{c}}$ and $\omega_{\mathrm{q}}$ and the white arrows the two resonant points $\pm\delta_{\mathrm{r}}$ at $\omega_{\mathrm{c}}=\omega_{\mathrm{q}}$. (c), The theoretical model of Eqs. (\ref{eq:reflCoefficient}) and (\ref{eq:reflAmplitude}) fitted to the experimental data of panel (b). (d), $|r|^{2}$ measured at the left resonant point in panel (b) indicated by a white arrow. The hybridized states are clearly resolved at $\omega/2\pi=\SI{6.4}{GHz}$ and $\omega/2\pi=\SI{7.0}{GHz}$. (e), $|r|^{2}$ measured at the right resonant point in panel (b).
    \label{fig2}
		}
\end{figure*}

\section{DEVICE CONFIGURATION}

Figure~\ref{fig1} shows our resonator-DQD device. The resonator is made of a series of Josephson junctions with a large inductance, yielding an impedance larger than $\SI{50}{\Omega}$, and therefore a larger coupling to the DQD~\cite{Stockklauser2017,Scarlino2022}.
The DQD is material-defined by InAs polytype structure~\cite{Nilsson2016,Barker2019,Junger2019,Junger2021,Ungerer2023}. In earlier studies, DQDs were formed in a two-dimensional electron gas like GaAs or Si by depletion gates~\cite{Stockklauser2017,Mi2017,Scarlino2022}. In contrast to these devices, our material-defined DQD requires minimal number of gate lines minimizing the charge noise from external sources. The gate voltages $V_{\mathrm{GL}}$ and $V_{\mathrm{GR}}$ shown in Fig.~\ref{fig1} (a) control the electron numbers of the quantum dots, whereas $V_{\mathrm{GM}}$ tunes the transparency of the interdot tunnel barrier. The bias voltage $V_{\mathrm{SD}}$ is applied to the source contact and the resulting current is measured at the drain contact. The DQD is coupled galvanically to a quarter-wavelength Josephson junction array resonator probed by RF reflectometry. All the electrical lines and the ground plane surrounding them are sputtered $\SI{100}{nm}$ thick niobium on a high-resistivity intrinsic silicon wafer with a $\SI{200}{nm}$ thermal silicon oxide coating. A $\SI{30}{nm}$ thick aluminum oxide is ALD-grown on top of the DC lines to act as an insulating layer between the lines and an additional ground plane of e-beam evaporated aluminum~\cite{Havir2023}. This ground plane increases the capacitance of the DC lines and thus reduces photon losses by increasing the impedance mismatch between the high-impedance resonator and the DC lines. Photons are fed into our quarter-wavelength resonator via a capacitive coupler at the voltage antinode on the right side of Fig.~\ref{fig1} (b). At the same antinode, the resonator is galvanically connected to the source contact of the DQD. The inset shows a few of the 110 Josephson junctions of our resonator that has a Lorentzian response at a resonance frequency of $\omega_{\mathrm{c}}/2\pi=\SI{6.7}{GHz}$ and a linewidth of $\kappa/2\pi=\SI{30}{MHz}$. The linewidth $\kappa$ is set by the sum of the resonator loss parameters as $\kappa=\kappa_{\mathrm{c}}+\kappa_{\mathrm{int}}$, where $\kappa_{\mathrm{c}}/2\pi=\SI{19}{MHz}$ is the capacitive input coupling between the resonator and the RF line and $\kappa_{\mathrm{int}}/2\pi=\SI{11}{MHz}$ is the internal loss rate. The DC junction resistance of $R_{\mathrm{T}}=\SI{500}{\Omega}$ is measured on the nominally identical test Josephson junctions in the top left corner of Fig.~\ref{fig1} (a). $\omega_{\mathrm{c}}$ and $R_{\mathrm{T}}$ allow us to estimate a characteristic impedance $Z_{0}\approx \SI{1.5}{k\Omega}$~\cite{Goppl2008,Ranni2023}. At the voltage node on the left side of Fig.~\ref{fig1} (b) the resonator is galvanically connected to a low-pass filtered DC voltage line to provide a bias voltage for the DQD~\cite{Haldar2023}. A zoom-in of the nanowire is shown in Fig.~\ref{fig1} (c). Our DQD is defined in a polytype InAs nanowire by zincblende dots formed between wurzite barriers~\cite{Barker2019,Khan2021,Haldar2023,Havir2023}. The inset displays a zoom-in of our nanowire before contact fabrication and the three barriers defining our DQD are pointed out by the white arrows. The measured reflection coefficient $|r|^{2}$ of the DQD-resonator hybrid is plotted in Fig.~\ref{fig1} (d) as a function of the two side gate voltages, exhibiting a very regular DQD honeycomb pattern. In this measurement $V_{\mathrm{GM}}=\SI{250}{mV}$ was fixed and we applied a RF drive at the cavity resonance frequency $\omega_{\mathrm{c}}$ with a power of $\SI{1}{aW}$. The white dashed hexagon indicates a fixed DQD charge state. The bright diagonal lines are so-called interdot transition lines. When crossing one of these lines an electron is shuttled from one dot to the other. The electric field of the resonator couples to the dipole moment of the DQD which becomes large at the DQD charge degeneracy points, where an electron can tunnel between the left and the right quantum dot~\cite{Frey2012,Petta2005,Stockklauser2017}. The reflection coefficient reaches almost total reflection $|r|^{2}=1$ at these lines as the resonator frequency is shifted by much more than $\kappa$ by the interaction with the DQD. The coupling strength $g$ between the resonator and the DQD is not the same for all interdot lines but depends on the charge configuration of the DQD in a non-trivial way as the many-electron wavefunction affects the interdot tunnel coupling $t$.

\begin{figure*}
    \centering
	\includegraphics[width=\textwidth]{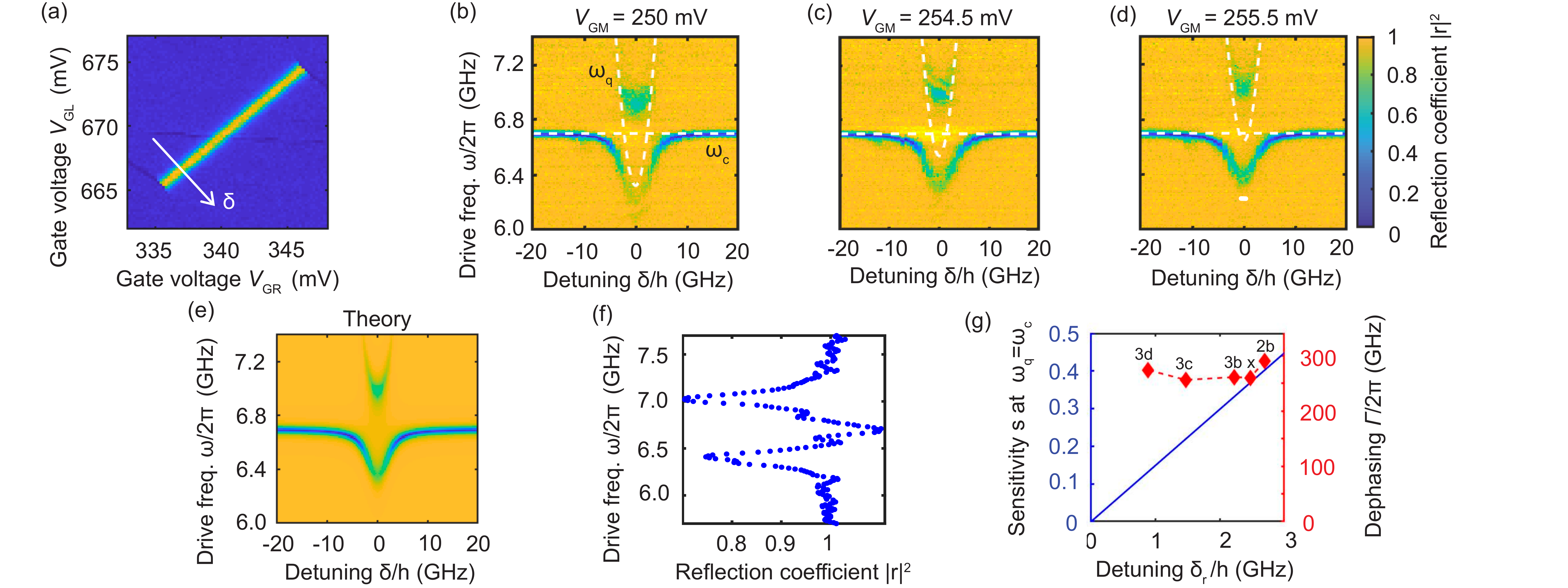}
	\caption{(a), An interdot line at fixed $V_{\mathrm{GM}}=\SI{250}{mV}$. The white arrow shows the detuning $\delta$ axis. (b), Measured $|r|^{2}$ as a function of the drive frequency $\omega$ and the detuning along the line shown in panel (a) at $V_{\mathrm{GM}}=\SI{250}{mV}$. The dashed white lines indicate $\omega_{\mathrm{c}}$ and $\omega_{\mathrm{q}}$. (c), Measured $|r|^{2}$ at $V_{\mathrm{GM}}=\SI{254.5}{mV}$. (d), Measured $|r|^{2}$ at $V_{\mathrm{GM}}=\SI{255.5}{mV}$. The small white line below the response at $\delta = 0$ shows the linewidth in detuning direction. (e), The theory model of Eqs. (\ref{eq:reflCoefficient}) and (\ref{eq:reflAmplitude}) fitted to the experimental data of panel (d). (f), $|r|^{2}$ measured at the resonant point $\delta=0$ of panel (d). (g), The dephasing rate $\Gamma$ and calculated sensitivity $s$ as a function of the detuning. The solid red diamonds are the measured values from  Figs.~\ref{fig3} (b) - (d) and Fig.~\ref{fig2} (b) as indicated with the labels. The diamond labeled with x is an additional data point from an interdot line at $V_{\mathrm{GL}}=\SI{370}{mV}$, $V_{\mathrm{GR}}=\SI{430}{mV}$ and $V_{\mathrm{GM}}=\SI{150}{mV}$. The dashed red line is a guide to the eye. The solid blue line is given by $s=\delta_{\mathrm{r}}/(\hbar\omega_{\mathrm{c}})$.  
	\label{fig3} 
		}
\end{figure*}

\section{CHARGE NOISE SENSITIVE OPERATION POINT}

\subsection{Strong coherent coupling}

Figure~\ref{fig2} (a) displays one interdot line measured as in Fig.~\ref{fig1} (d) at $V_{\mathrm{GM}}=\SI{150}{mV}$. Figure~\ref{fig2} (b) shows $|r| ^{2}$ measured as a function of the drive frequency $\omega$ and the gate voltages $V_{\mathrm{GL}}$ and $V_{\mathrm{GR}}$ along the detuning $\delta$ axis shown in panel (a). $\delta$ is the difference between the left and the right quantum dot energy levels. $\delta$ and $t$ set the excitation energy of the DQD, $E_{\mathrm{DQD}}=\sqrt{\delta^{2}+(2t)^{2}}$~\cite{Childress2004}. In Fig.~\ref{fig2} (b) at large $|\delta|$ we see only the resonance of the bare cavity at $\omega_{\mathrm{c}}/2\pi=\SI{6.7}{GHz}$. In this regime the qubit frequency $\omega_{\mathrm{q}}=E_{\mathrm{DQD}}/\hbar$ is considerably larger than $\omega_{\mathrm{c}}$ as clearly shown by the white dashed lines denoting the qubit and the cavity. Towards lower $|\delta|$ the response starts to deviate from the bare cavity case. The qubit interacts with the cavity as $\omega_{\mathrm{q}}$ approaches $\omega_{\mathrm{c}}$. The two white arrows in Fig.~\ref{fig2} (b) indicate the resonant finite detuning points where the corresponding avoided crossings develop around the points $\omega_{\mathrm{q}}=\omega_{\mathrm{c}}$. The two avoided crossings are manifestations of strong coupling $g$ between the resonator and the qubit hybridizing the photonic and the electronic states and splitting the states in energy. The two hybridized states appear particularly pronounced in the vertical linecuts of Figs.~\ref{fig2} (d) and (e) measured at the negative and the positive resonant detuning points indicated in Fig.~\ref{fig2} (b).     

To model the reflection coefficient response, an input-output theory together with a Jaynes-Cummings Hamiltonian modelling the resonator-DQD system is employed (see Appendix A). Under the assumption of small drive power, we obtain the response analytically as
\begin{equation}
    |r(\omega)|^{2}=|1-\kappa_{\mathrm{c}}A(\omega)|^{2},
    \label{eq:reflCoefficient}
\end{equation}
with
\begin{equation}
    A(\omega)=\frac{\Gamma/2-i(\omega-\omega_{\mathrm{q}})}{(\kappa/2-i(\omega-\omega_{\mathrm{c}}))(\Gamma/2-i(\omega-\omega_{\mathrm{q}}))+g^{2}},
    \label{eq:reflAmplitude}
\end{equation}
where $\Gamma$ stands for the total dephasing rate of the DQD. Figure~\ref{fig2} (c) shows the calculated reflection coefficient of Eq.~(\ref{eq:reflCoefficient}) fitted to the experimental response in Fig.~\ref{fig2} (b). The fit yields $g/2\pi=\SI{270}{MHz}$, $\Gamma/2\pi=\SI{290}{MHz}$, $t/h=\SI{3.1}{GHz}$ and the right gate lever arm $\alpha=\SI{120}{MHz/GHz}$ as well as the resonant detuning points $\delta_{\mathrm{r}}/h=\pm\SI{2.7}{GHz}$ with $\omega_{\mathrm{c}}$, $\kappa_{\mathrm{c}}$ and $\kappa_{\mathrm{int}}$ fixed to the aforementioned values. These values bring the hybrid system into the strong coupling regime, with $2g/(\Gamma+\kappa/2)=1.8>1$~\cite{Blais2021}. Note that the dephasing of the DQD only affects the reflection coefficient at detunings close to resonance. From Fig.~\ref{fig2} (b), we thus extract $\Gamma$ at $|\delta|$ corresponding to the white arrows. The theoretical model describes essentially all the features of the experiment quantitatively correct. Importantly, this procedure allows us to convert the gate voltage detuning axis into an energy. In Figs.~\ref{fig2} (b) and (c) the right gate voltage change $\Delta V_{\mathrm{GR}}$ is converted to detuning as $\delta=e\alpha \Delta V_{\mathrm{GR}}$.

\subsection{Sensitivity to charge noise}

Ambient charge noise couples to the energy levels of the quantum dots and hence perturbs the DQD energy increasing the total dephasing rate $\Gamma$. Possible sources for charge noise are for instance voltage fluctuations in the gate and bias lines, and charge fluctuations in the nanowire or its vicinity, e.g. in the substrate surface~\cite{Dial2013,Bermeister2014}. Our material-defined DQD requires minimal amount of gate lines which provides an ideal system to obtain low gate noise. A typical way to minimize the effect of the charge noise is to operate the system in a charge insensitive point~\cite{Petersson2010,Vion2002,Mi2017,Scarlino2022}. We quantify the DQD sensitivity to charge noise by the slope of the DQD energy as $s=dE_{\mathrm{DQD}}/d\delta=1/\sqrt{1+(2t/\delta)^{2}}$. The sensitivity at the resonant points denoted by the white arrows in Fig.~\ref{fig2} (b) is $s=0.39$. These finite detuning points are sensitive to charge noise as fluctuations in $\delta$ change the qubit frequency. Towards larger detuning, the sensitivity approaches the limit of $s=1$ while the lower limit is $s=0$ at $\delta=0$. At the latter point a small change in $\delta$ does not significantly affect the qubit frequency. To tune the qubit at this point, the interdot tunnel coupling $t$ has to be adjusted such that the resonant condition $\omega_{\mathrm{q}}=\omega_{\mathrm{c}}$ holds at $\delta=0$. This constellation is often called "sweet spot"~\cite{Nakamura1999}. At sweet spot the qubit is first-order robust against charge noise and hence it is of significant interest for charge qubit community and a subject for several studies~\cite{Vion2002,Petersson2010,Medford2013,Mi2017,Thorgrimsson2017,Scarlino2022}. 

\section{CHARGE NOISE INSENSITIVE OPERATION POINT (SWEET SPOT)}

\subsection{Strong coherent coupling}
 
As a next step, we now show similar experiments as above, for configurations closer and closer to the sweet spot. Figures~\ref{fig3} (a) and (b) show the same type of measurement for a second interdot transition with the middle gate voltage increased to $V_{\mathrm{GM}}=\SI{250}{mV}$ to reduce the interdot barrier strength. This increases $t$ and brings the DQD closer to the sweet spot. In Fig.~\ref{fig3} (b) we now find both branches at $\delta=0$ clearly closer to the sweet spot than in Fig.~\ref{fig2} (b) and $\omega_{\mathrm{q}}=\omega_{\mathrm{c}}$ at smaller $|\delta|$. However, Fig.~\ref{fig3} (b) is still slightly off the sweet spot as the two hybridized states are located at non-equal energy differences from the bare resonator frequency and the higher energy resonance shows a stronger response. Monitoring the same interdot transition, in Fig.~\ref{fig3} (c) we further increased the middle gate voltage to $V_{\mathrm{GM}}=\SI{254.5}{mV}$ to tune closer to the sweet spot. Now the response of the two states at $\delta=0$ is more equal than in panel (b) and also the distances in energy from $\omega_{\mathrm{c}}$ are closer to each other. By further increasing the middle gate voltage to $V_{\mathrm{GM}}=\SI{255.5}{mV}$ in Fig.~\ref{fig3} (d) we get even closer to the sweet spot. $|r|^{2}$ in Fig.~\ref{fig3} (f) is recorded at $\delta=0$ and shows two well-resolved resonances with equally strong responses and distances from $\omega_{\mathrm{c}}$. Since the reflection is only affected by $\Gamma$ when the DQD is close to resonance with the cavity, fitting Fig.~\ref{fig3} (d) allows us to extract $\Gamma$ at the sweet spot. Figure~\ref{fig3} (e) shows our theoretical model fitted to the experimental data in panel (d). Again, the theory matches well with the experimental data and yields $g/2\pi=\SI{320}{MHz}$, $\Gamma/2\pi=\SI{260}{MHz}$, $t/h=\SI{3.4}{GHz}$ and the right gate lever arm $\alpha=\SI{130}{MHz/GHz}$. The right gate lever arm we obtain from a finite bias triangle measurement, $\SI{150}{MHz/GHz}$, is close to the values of the fits in Figs.~\ref{fig2} (c) and \ref{fig3} (e).

\subsection{Sensitivity to charge noise}

To determine the sensitivity to charge noise at the sweet spot of Fig.~\ref{fig3} (d), we estimate the fluctuations in the detuning to be $\Delta\delta_{\pm}/h=\pm\SI{0.5}{GHz}$ or smaller. This estimate is highlighted in Fig.~\ref{fig3} (d) by the small white line below the response at $\delta=0$ and it is based on the linewidth of the features in the $\delta$-direction. The estimate together with the fitted $t$ yield an upper limit of the sensitivity $s=0.08$ at the sweet spot. Hence the sensitivity at the sweet spot is at least a factor of 5 smaller than in the finite detuning case. The total dephasing rates $\Gamma$ at these two operation points however differ only by $\SI{10}{\%}$. Therefore the charge noise cannot be the dominating dephasing mechanism in our device. Figure~\ref{fig3} (g) summarizes the results of Figs.~\ref{fig2} and \ref{fig3} by visualizing that $s$ (in blue) and $\Gamma$ (in red) have different dependencies on detuning. $\Gamma$ remains constant within $\SI{10}{\%}$ whereas $s$ increases linearly as a function of $\delta_{\mathrm{r}}$. The solid blue lines is $s=\delta_{\mathrm{r}}/(\hbar\omega_{\mathrm{c}})$ obtained by inserting $\omega_{\mathrm{q}}=\omega_{\mathrm{c}}$ into $s=1/\sqrt{1+(2t/\delta_{\mathrm{r}})^{2}}$. 

\section{DISCUSSION}

In addition to charge noise, decoherence of charge qubits arises from electron-phonon relaxation of the DQD~\cite{fujisawa1998, hartke2018, Maisi2020} as well as dissipation and radiative losses in the DQD leads~\cite{Mi2017b, guthrie2022}. We estimate that the electron-phonon relaxation via the piezo-electric coupling is likely the dominant source of dephasing in our system with the following arguments: The electron-phonon relaxation rate study of Ref.~\citealp{Maisi2020} estimates that this relaxation rate is about $\SI{100}{MHz}$ in a GaAs based DQD qubit. We anticipate that our InAs DQD has a similar rate as a similar type of piezoelectric coupling is present in both materials. The heavier In atoms as compared to the Ga atoms increase the piezoelectric coupling, but also precise geometry of the DQD states and the phonon spectrum may influence on the precise value of rate. The above estimate matches well with measured dephasing rates in GaAs quantum dots~\cite{Stockklauser2017, Scarlino2022} as well as the value determined from our data. Also interestingly, the Si DQDs reach more than an order of magnitude lower dephasing rates~\cite{Mi2017}. This fits the picture too as silicon does not have the piezoelectric coupling. All of these evidences therefore suggest that the piezoelectric electron-phonon relaxation would be the dominant mechanism in our device. In addition, the resistive and radiative losses in the leads connecting to the DQD may contribute to the dephasing. We have partially mitigated these effects with the added capacitance to the lines. Further experiments beyond the scope of this study, for example by using fully superconducting contact lines and variations in the line impedance design, would be needed to obtain a definite answer to these considerations. Other interesting avenues for future studies to test and minimize the phonon contribution would be to orient the DQD axis to the crystalline direction with suppressed electron-phonon interaction~\cite{Golovach2008}, or take use of the destructive interference effect in electron-phonon coupling in one dimensional phonon systems~\cite{Roulleau2011}.

\section{CONCLUSION}

In conclusion, we have realized a strong coherent coupling between a high-impedance resonator and a material-defined DQD in an InAs polytype nanowire. We showed that the charge noise is not the dominant source of decoherence and discussed that the piezoelectric electron-phonon relaxation in the DQD is the most likely origin of the observed dephasing rate. These results therefore present a charge based qubit where the coherence is not limited by charge fluctuations as is typically the case. 

\section*{ACKNOWLEDGMENTS}

We acknowledge Peter Samuelsson, Jann Hinnerk Ungerer, Alessia Pally and Artem Kononov for fruitful discussions and Swedish Research Council (Dnr 2019-04111), the Foundational Questions Institute, a donor advised fund of Silicon Valley Community Foundation (grant number FQXi-IAF19-07), the Swiss National Science Foundation (Eccellenza Professorial Fellowship PCEFP2\_194268), Knut and Alice Wallenberg Foundation through the Wallenberg Centre for Quantum Technology (WACQT) and NanoLund for financial support. CS acknowledges support from the SNSF through grant 192027 and the NCCR-Spin. PS acknowledges support from the SNSF through grant 200418 and the SERI through grant 589025.

\appendix

\section{REFLECTION COEFFICIENT}
\label{app:reflection}
We model the resonator-DQD hybrid device by the Jaynes-Cummings Hamiltonian ($\hbar=1$)
\begin{equation}
\label{eq:hamiltonJC}
\hat{H} = \omega_\mathrm{c}\hat{a}^\dagger\hat{a}+\frac{\omega_\mathrm{q}}{2}\hat{\sigma}_\mathrm{z}+g\left(\hat{a}\hat{\sigma}^\dagger+\hat{a}^\dagger\hat{\sigma}\right),
\end{equation}
where $\hat{a}$ is the photon annihilation operator, $\sigma$ the qubit lowering operator, and $\sigma_\mathrm{z}$ the Pauli z-matrix in the qubit subspace.
To derive the response of the resonator, we use the equations of motion \cite{gardiner_book}
\begin{equation}
	\label{eq:ainoutgen}
	\begin{aligned}
		&\partial_t{\langle \hat{a}\rangle}(t) = -i\omega_\mathrm{c} \hat{a}(t)-ig\langle \hat\sigma\rangle (t)-\frac{\kappa}{2}\langle \hat{a}\rangle(t)-\sqrt{\kappa_{\mathrm c}}\langle \hat{b}_{\rm in}\rangle(t),\\&
		\partial_t \langle \hat{\sigma}\rangle (t) = -i\omega_\mathrm{q}\langle \hat{\sigma}\rangle (t)+ig\langle \hat{a} \hat{\sigma}_\mathrm{z}\rangle(t)-\frac{{\Gamma}}{2}\langle \hat{\sigma}\rangle (t),
	\end{aligned}
\end{equation}
where $\hat{b}_{\rm in}(t)$ describes the coherent drive. In the low-drive limit, the qubit remains approximately in the ground-state and we may replace $\langle\hat{a} \hat{\sigma}_\mathrm{z}\rangle(t)\rightarrow -\langle\hat{a}\rangle(t)$.
The reflected output of the cavity can be computed from the input-output relation \cite{gardiner_book}
\begin{equation}
	\label{eq:inout}
	\langle \hat{b} _{\rm out}\rangle(t) = \langle \hat{b}_{\rm in}\rangle(t)+ \sqrt{\kappa_{\mathrm{c}}}\langle \hat{a}\rangle(t).
\end{equation}
Upon Fourier transformation, Eqs.~\eqref{eq:ainoutgen} and \eqref{eq:inout} may be solved resulting in the reflection amplitude $r(\omega) = \langle \hat{b} _{\rm out}\rangle(\omega)/\langle \hat{b} _{\rm in}\rangle(\omega)$ given in Eqs.~\eqref{eq:reflCoefficient} and \eqref{eq:reflAmplitude} in the main text.

\bibliography{main}

\begin{thebibliography}{45}%
\makeatletter
\providecommand \@ifxundefined [1]{%
 \@ifx{#1\undefined}
}%
\providecommand \@ifnum [1]{%
 \ifnum #1\expandafter \@firstoftwo
 \else \expandafter \@secondoftwo
 \fi
}%
\providecommand \@ifx [1]{%
 \ifx #1\expandafter \@firstoftwo
 \else \expandafter \@secondoftwo
 \fi
}%
\providecommand \natexlab [1]{#1}%
\providecommand \enquote  [1]{``#1''}%
\providecommand \bibnamefont  [1]{#1}%
\providecommand \bibfnamefont [1]{#1}%
\providecommand \citenamefont [1]{#1}%
\providecommand \href@noop [0]{\@secondoftwo}%
\providecommand \href [0]{\begingroup \@sanitize@url \@href}%
\providecommand \@href[1]{\@@startlink{#1}\@@href}%
\providecommand \@@href[1]{\endgroup#1\@@endlink}%
\providecommand \@sanitize@url [0]{\catcode `\\12\catcode `\$12\catcode
  `\&12\catcode `\#12\catcode `\^12\catcode `\_12\catcode `\%12\relax}%
\providecommand \@@startlink[1]{}%
\providecommand \@@endlink[0]{}%
\providecommand \url  [0]{\begingroup\@sanitize@url \@url }%
\providecommand \@url [1]{\endgroup\@href {#1}{\urlprefix }}%
\providecommand \urlprefix  [0]{URL }%
\providecommand \Eprint [0]{\href }%
\providecommand \doibase [0]{http://dx.doi.org/}%
\providecommand \selectlanguage [0]{\@gobble}%
\providecommand \bibinfo  [0]{\@secondoftwo}%
\providecommand \bibfield  [0]{\@secondoftwo}%
\providecommand \translation [1]{[#1]}%
\providecommand \BibitemOpen [0]{}%
\providecommand \bibitemStop [0]{}%
\providecommand \bibitemNoStop [0]{.\EOS\space}%
\providecommand \EOS [0]{\spacefactor3000\relax}%
\providecommand \BibitemShut  [1]{\csname bibitem#1\endcsname}%
\let\auto@bib@innerbib\@empty
%</preamble>
\bibitem [{\citenamefont {Haroche}\ and\ \citenamefont
  {Raimond}(2006)}]{Haroche2006}%
  \BibitemOpen
  \bibfield  {author} {\bibinfo {author} {\bibfnamefont {S.}~\bibnamefont
  {Haroche}}\ and\ \bibinfo {author} {\bibfnamefont {J.-M.}\ \bibnamefont
  {Raimond}},\ }\href {\doibase 10.1093/acprof:oso/9780198509141.001.0001}
  {\emph {\bibinfo {title} {{{Exploring the Quantum: Atoms, Cavities, and
  Photons}}}}}\ (\bibinfo  {publisher} {Oxford University Press},\ \bibinfo
  {year} {2006})\BibitemShut {NoStop}%
\bibitem [{\citenamefont {Thompson}\ \emph {et~al.}(1992)\citenamefont
  {Thompson}, \citenamefont {Rempe},\ and\ \citenamefont
  {Kimble}}]{Thompson1992}%
  \BibitemOpen
  \bibfield  {author} {\bibinfo {author} {\bibfnamefont {R.~J.}\ \bibnamefont
  {Thompson}}, \bibinfo {author} {\bibfnamefont {G.}~\bibnamefont {Rempe}}, \
  and\ \bibinfo {author} {\bibfnamefont {H.~J.}\ \bibnamefont {Kimble}},\
  }\bibfield  {title} {\enquote {\bibinfo {title} {{Observation of normal-mode
  splitting for an atom in an optical cavity}},}\ }\href {\doibase
  10.1103/PhysRevLett.68.1132} {\bibfield  {journal} {\bibinfo  {journal}
  {Phys. Rev. Lett.}\ }\textbf {\bibinfo {volume} {68}},\ \bibinfo {pages}
  {1132--1135} (\bibinfo {year} {1992})}\BibitemShut {NoStop}%
\bibitem [{\citenamefont {Brune}\ \emph {et~al.}(1996)\citenamefont {Brune},
  \citenamefont {Schmidt-Kaler}, \citenamefont {Maali}, \citenamefont {Dreyer},
  \citenamefont {Hagley}, \citenamefont {Raimond},\ and\ \citenamefont
  {Haroche}}]{Brune1996}%
  \BibitemOpen
  \bibfield  {author} {\bibinfo {author} {\bibfnamefont {M.}~\bibnamefont
  {Brune}}, \bibinfo {author} {\bibfnamefont {F.}~\bibnamefont
  {Schmidt-Kaler}}, \bibinfo {author} {\bibfnamefont {A.}~\bibnamefont
  {Maali}}, \bibinfo {author} {\bibfnamefont {J.}~\bibnamefont {Dreyer}},
  \bibinfo {author} {\bibfnamefont {E.}~\bibnamefont {Hagley}}, \bibinfo
  {author} {\bibfnamefont {J.~M.}\ \bibnamefont {Raimond}}, \ and\ \bibinfo
  {author} {\bibfnamefont {S.}~\bibnamefont {Haroche}},\ }\bibfield  {title}
  {\enquote {\bibinfo {title} {{Quantum Rabi Oscillation: A Direct Test of
  Field Quantization in a Cavity}},}\ }\href {\doibase
  10.1103/PhysRevLett.76.1800} {\bibfield  {journal} {\bibinfo  {journal}
  {Phys. Rev. Lett.}\ }\textbf {\bibinfo {volume} {76}},\ \bibinfo {pages}
  {1800--1803} (\bibinfo {year} {1996})}\BibitemShut {NoStop}%
\bibitem [{\citenamefont {Wallraff}\ \emph {et~al.}(2004)\citenamefont
  {Wallraff}, \citenamefont {Schuster}, \citenamefont {Blais}, \citenamefont
  {Frunzio}, \citenamefont {Huang}, \citenamefont {Majer}, \citenamefont
  {Kumar}, \citenamefont {Girvin},\ and\ \citenamefont
  {Schoelkopf}}]{Wallraff2004}%
  \BibitemOpen
  \bibfield  {author} {\bibinfo {author} {\bibfnamefont {A.}~\bibnamefont
  {Wallraff}}, \bibinfo {author} {\bibfnamefont {D.~I.}\ \bibnamefont
  {Schuster}}, \bibinfo {author} {\bibfnamefont {A.}~\bibnamefont {Blais}},
  \bibinfo {author} {\bibfnamefont {L.}~\bibnamefont {Frunzio}}, \bibinfo
  {author} {\bibfnamefont {R.-S.}\ \bibnamefont {Huang}}, \bibinfo {author}
  {\bibfnamefont {J.}~\bibnamefont {Majer}}, \bibinfo {author} {\bibfnamefont
  {S.}~\bibnamefont {Kumar}}, \bibinfo {author} {\bibfnamefont {S.~M.}\
  \bibnamefont {Girvin}}, \ and\ \bibinfo {author} {\bibfnamefont {R.~J.}\
  \bibnamefont {Schoelkopf}},\ }\bibfield  {title} {\enquote {\bibinfo {title}
  {{Strong coupling of a single photon to a superconducting qubit using circuit
  quantum electrodynamics}},}\ }\href {\doibase 10.1038/nature02851} {\bibfield
   {journal} {\bibinfo  {journal} {Nature}\ }\textbf {\bibinfo {volume}
  {431}},\ \bibinfo {pages} {162--167} (\bibinfo {year} {2004})}\BibitemShut
  {NoStop}%
\bibitem [{\citenamefont {Reithmaier}\ \emph {et~al.}(2004)\citenamefont
  {Reithmaier}, \citenamefont {S{\k{e}}k}, \citenamefont {L{\"o}ffler},
  \citenamefont {Hofmann}, \citenamefont {Kuhn}, \citenamefont {Reitzenstein},
  \citenamefont {Keldysh}, \citenamefont {Kulakovskii}, \citenamefont
  {Reinecke},\ and\ \citenamefont {Forchel}}]{Reithmaier2004}%
  \BibitemOpen
  \bibfield  {author} {\bibinfo {author} {\bibfnamefont {J.~P.}\ \bibnamefont
  {Reithmaier}}, \bibinfo {author} {\bibfnamefont {G.}~\bibnamefont
  {S{\k{e}}k}}, \bibinfo {author} {\bibfnamefont {A.}~\bibnamefont
  {L{\"o}ffler}}, \bibinfo {author} {\bibfnamefont {C.}~\bibnamefont
  {Hofmann}}, \bibinfo {author} {\bibfnamefont {S.}~\bibnamefont {Kuhn}},
  \bibinfo {author} {\bibfnamefont {S.}~\bibnamefont {Reitzenstein}}, \bibinfo
  {author} {\bibfnamefont {L.~V.}\ \bibnamefont {Keldysh}}, \bibinfo {author}
  {\bibfnamefont {V.~D.}\ \bibnamefont {Kulakovskii}}, \bibinfo {author}
  {\bibfnamefont {T.~L.}\ \bibnamefont {Reinecke}}, \ and\ \bibinfo {author}
  {\bibfnamefont {A.}~\bibnamefont {Forchel}},\ }\bibfield  {title} {\enquote
  {\bibinfo {title} {{Strong coupling in a single quantum dot--semiconductor
  microcavity system}},}\ }\href {\doibase 10.1038/nature02969} {\bibfield
  {journal} {\bibinfo  {journal} {Nature}\ }\textbf {\bibinfo {volume} {432}},\
  \bibinfo {pages} {197--200} (\bibinfo {year} {2004})}\BibitemShut {NoStop}%
\bibitem [{\citenamefont {Yoshie}\ \emph {et~al.}(2004)\citenamefont {Yoshie},
  \citenamefont {Scherer}, \citenamefont {Hendrickson}, \citenamefont
  {Khitrova}, \citenamefont {Gibbs}, \citenamefont {Rupper}, \citenamefont
  {Ell}, \citenamefont {Shchekin},\ and\ \citenamefont {Deppe}}]{Yoshie2004}%
  \BibitemOpen
  \bibfield  {author} {\bibinfo {author} {\bibfnamefont {T.}~\bibnamefont
  {Yoshie}}, \bibinfo {author} {\bibfnamefont {A.}~\bibnamefont {Scherer}},
  \bibinfo {author} {\bibfnamefont {J.}~\bibnamefont {Hendrickson}}, \bibinfo
  {author} {\bibfnamefont {G.}~\bibnamefont {Khitrova}}, \bibinfo {author}
  {\bibfnamefont {H.~M.}\ \bibnamefont {Gibbs}}, \bibinfo {author}
  {\bibfnamefont {G.}~\bibnamefont {Rupper}}, \bibinfo {author} {\bibfnamefont
  {C.}~\bibnamefont {Ell}}, \bibinfo {author} {\bibfnamefont {O.~B.}\
  \bibnamefont {Shchekin}}, \ and\ \bibinfo {author} {\bibfnamefont {D.~G.}\
  \bibnamefont {Deppe}},\ }\bibfield  {title} {\enquote {\bibinfo {title}
  {{Vacuum Rabi splitting with a single quantum dot in a photonic crystal
  nanocavity}},}\ }\href {\doibase 10.1038/nature03119} {\bibfield  {journal}
  {\bibinfo  {journal} {Nature}\ }\textbf {\bibinfo {volume} {432}},\ \bibinfo
  {pages} {200--203} (\bibinfo {year} {2004})}\BibitemShut {NoStop}%
\bibitem [{\citenamefont {Majer}\ \emph {et~al.}(2007)\citenamefont {Majer},
  \citenamefont {Chow}, \citenamefont {Gambetta}, \citenamefont {Koch},
  \citenamefont {Johnson}, \citenamefont {Schreier}, \citenamefont {Frunzio},
  \citenamefont {Schuster}, \citenamefont {Houck}, \citenamefont {Wallraff},
  \citenamefont {Blais}, \citenamefont {Devoret}, \citenamefont {Girvin},\ and\
  \citenamefont {Schoelkopf}}]{Majer2007}%
  \BibitemOpen
  \bibfield  {author} {\bibinfo {author} {\bibfnamefont {J.}~\bibnamefont
  {Majer}}, \bibinfo {author} {\bibfnamefont {J.~M.}\ \bibnamefont {Chow}},
  \bibinfo {author} {\bibfnamefont {J.~M.}\ \bibnamefont {Gambetta}}, \bibinfo
  {author} {\bibfnamefont {Jens}\ \bibnamefont {Koch}}, \bibinfo {author}
  {\bibfnamefont {B.~R.}\ \bibnamefont {Johnson}}, \bibinfo {author}
  {\bibfnamefont {J.~A.}\ \bibnamefont {Schreier}}, \bibinfo {author}
  {\bibfnamefont {L.}~\bibnamefont {Frunzio}}, \bibinfo {author} {\bibfnamefont
  {D.~I.}\ \bibnamefont {Schuster}}, \bibinfo {author} {\bibfnamefont {A.~A.}\
  \bibnamefont {Houck}}, \bibinfo {author} {\bibfnamefont {A.}~\bibnamefont
  {Wallraff}}, \bibinfo {author} {\bibfnamefont {A.}~\bibnamefont {Blais}},
  \bibinfo {author} {\bibfnamefont {M.~H.}\ \bibnamefont {Devoret}}, \bibinfo
  {author} {\bibfnamefont {S.~M.}\ \bibnamefont {Girvin}}, \ and\ \bibinfo
  {author} {\bibfnamefont {R.~J.}\ \bibnamefont {Schoelkopf}},\ }\bibfield
  {title} {\enquote {\bibinfo {title} {{Coupling superconducting qubits via a
  cavity bus}},}\ }\href {\doibase 10.1038/nature06184} {\bibfield  {journal}
  {\bibinfo  {journal} {Nature}\ }\textbf {\bibinfo {volume} {449}},\ \bibinfo
  {pages} {443--447} (\bibinfo {year} {2007})}\BibitemShut {NoStop}%
\bibitem [{\citenamefont {Sillanp{\"a}{\"a}}\ \emph {et~al.}(2007)\citenamefont
  {Sillanp{\"a}{\"a}}, \citenamefont {Park},\ and\ \citenamefont
  {Simmonds}}]{Sillanpaa2007}%
  \BibitemOpen
  \bibfield  {author} {\bibinfo {author} {\bibfnamefont {M.~A.}\ \bibnamefont
  {Sillanp{\"a}{\"a}}}, \bibinfo {author} {\bibfnamefont {J.~I.}\ \bibnamefont
  {Park}}, \ and\ \bibinfo {author} {\bibfnamefont {R.~W.}\ \bibnamefont
  {Simmonds}},\ }\bibfield  {title} {\enquote {\bibinfo {title} {{Coherent
  quantum state storage and transfer between two phase qubits via a resonant
  cavity}},}\ }\href {\doibase 10.1038/nature06124} {\bibfield  {journal}
  {\bibinfo  {journal} {Nature}\ }\textbf {\bibinfo {volume} {449}},\ \bibinfo
  {pages} {438--442} (\bibinfo {year} {2007})}\BibitemShut {NoStop}%
\bibitem [{\citenamefont {Houck}\ \emph {et~al.}(2007)\citenamefont {Houck},
  \citenamefont {Schuster}, \citenamefont {Gambetta}, \citenamefont {Schreier},
  \citenamefont {Johnson}, \citenamefont {Chow}, \citenamefont {Frunzio},
  \citenamefont {Majer}, \citenamefont {Devoret}, \citenamefont {Girvin},\ and\
  \citenamefont {Schoelkopf}}]{Houck2007}%
  \BibitemOpen
  \bibfield  {author} {\bibinfo {author} {\bibfnamefont {A.~A.}\ \bibnamefont
  {Houck}}, \bibinfo {author} {\bibfnamefont {D.~I.}\ \bibnamefont {Schuster}},
  \bibinfo {author} {\bibfnamefont {J.~M.}\ \bibnamefont {Gambetta}}, \bibinfo
  {author} {\bibfnamefont {J.~A.}\ \bibnamefont {Schreier}}, \bibinfo {author}
  {\bibfnamefont {B.~R.}\ \bibnamefont {Johnson}}, \bibinfo {author}
  {\bibfnamefont {J.~M.}\ \bibnamefont {Chow}}, \bibinfo {author}
  {\bibfnamefont {L.}~\bibnamefont {Frunzio}}, \bibinfo {author} {\bibfnamefont
  {J.}~\bibnamefont {Majer}}, \bibinfo {author} {\bibfnamefont {M.~H.}\
  \bibnamefont {Devoret}}, \bibinfo {author} {\bibfnamefont {S.~M.}\
  \bibnamefont {Girvin}}, \ and\ \bibinfo {author} {\bibfnamefont {R.~J.}\
  \bibnamefont {Schoelkopf}},\ }\bibfield  {title} {\enquote {\bibinfo {title}
  {{Generating single microwave photons in a circuit}},}\ }\href {\doibase
  10.1038/nature06126} {\bibfield  {journal} {\bibinfo  {journal} {Nature}\
  }\textbf {\bibinfo {volume} {449}},\ \bibinfo {pages} {328--331} (\bibinfo
  {year} {2007})}\BibitemShut {NoStop}%
\bibitem [{\citenamefont {Eichler}\ \emph {et~al.}(2012)\citenamefont
  {Eichler}, \citenamefont {Lang}, \citenamefont {Fink}, \citenamefont
  {Govenius}, \citenamefont {Filipp},\ and\ \citenamefont
  {Wallraff}}]{Eichler2012}%
  \BibitemOpen
  \bibfield  {author} {\bibinfo {author} {\bibfnamefont {C.}~\bibnamefont
  {Eichler}}, \bibinfo {author} {\bibfnamefont {C.}~\bibnamefont {Lang}},
  \bibinfo {author} {\bibfnamefont {J.~M.}\ \bibnamefont {Fink}}, \bibinfo
  {author} {\bibfnamefont {J.}~\bibnamefont {Govenius}}, \bibinfo {author}
  {\bibfnamefont {S.}~\bibnamefont {Filipp}}, \ and\ \bibinfo {author}
  {\bibfnamefont {A.}~\bibnamefont {Wallraff}},\ }\bibfield  {title} {\enquote
  {\bibinfo {title} {{Observation of Entanglement between Itinerant Microwave
  Photons and a Superconducting Qubit}},}\ }\href {\doibase
  10.1103/PhysRevLett.109.240501} {\bibfield  {journal} {\bibinfo  {journal}
  {Phys. Rev. Lett.}\ }\textbf {\bibinfo {volume} {109}},\ \bibinfo {pages}
  {240501} (\bibinfo {year} {2012})}\BibitemShut {NoStop}%
\bibitem [{\citenamefont {Stockklauser}\ \emph {et~al.}(2017)\citenamefont
  {Stockklauser}, \citenamefont {Scarlino}, \citenamefont {Koski},
  \citenamefont {Gasparinetti}, \citenamefont {Andersen}, \citenamefont
  {Reichl}, \citenamefont {Wegscheider}, \citenamefont {Ihn}, \citenamefont
  {Ensslin},\ and\ \citenamefont {Wallraff}}]{Stockklauser2017}%
  \BibitemOpen
  \bibfield  {author} {\bibinfo {author} {\bibfnamefont {A.}~\bibnamefont
  {Stockklauser}}, \bibinfo {author} {\bibfnamefont {P.}~\bibnamefont
  {Scarlino}}, \bibinfo {author} {\bibfnamefont {J.~V.}\ \bibnamefont {Koski}},
  \bibinfo {author} {\bibfnamefont {S.}~\bibnamefont {Gasparinetti}}, \bibinfo
  {author} {\bibfnamefont {C.~K.}\ \bibnamefont {Andersen}}, \bibinfo {author}
  {\bibfnamefont {C.}~\bibnamefont {Reichl}}, \bibinfo {author} {\bibfnamefont
  {W.}~\bibnamefont {Wegscheider}}, \bibinfo {author} {\bibfnamefont
  {T.}~\bibnamefont {Ihn}}, \bibinfo {author} {\bibfnamefont {K.}~\bibnamefont
  {Ensslin}}, \ and\ \bibinfo {author} {\bibfnamefont {A.}~\bibnamefont
  {Wallraff}},\ }\bibfield  {title} {\enquote {\bibinfo {title} {{Strong
  Coupling Cavity QED with Gate-Defined Double Quantum Dots Enabled by a High
  Impedance Resonator}},}\ }\href {\doibase 10.1103/PhysRevX.7.011030}
  {\bibfield  {journal} {\bibinfo  {journal} {Phys. Rev. X}\ }\textbf {\bibinfo
  {volume} {7}},\ \bibinfo {pages} {011030} (\bibinfo {year}
  {2017})}\BibitemShut {NoStop}%
\bibitem [{\citenamefont {Mi}\ \emph {et~al.}(2017{\natexlab{a}})\citenamefont
  {Mi}, \citenamefont {Cady}, \citenamefont {Zajac}, \citenamefont {Deelman},\
  and\ \citenamefont {Petta}}]{Mi2017}%
  \BibitemOpen
  \bibfield  {author} {\bibinfo {author} {\bibfnamefont {X.}~\bibnamefont
  {Mi}}, \bibinfo {author} {\bibfnamefont {J.~V.}\ \bibnamefont {Cady}},
  \bibinfo {author} {\bibfnamefont {D.~M.}\ \bibnamefont {Zajac}}, \bibinfo
  {author} {\bibfnamefont {P.~W.}\ \bibnamefont {Deelman}}, \ and\ \bibinfo
  {author} {\bibfnamefont {J.~R.}\ \bibnamefont {Petta}},\ }\bibfield  {title}
  {\enquote {\bibinfo {title} {{Strong coupling of a single electron in silicon
  to a microwave photon}},}\ }\href {\doibase 10.1126/science.aal2469}
  {\bibfield  {journal} {\bibinfo  {journal} {Science}\ }\textbf {\bibinfo
  {volume} {355}},\ \bibinfo {pages} {156--158} (\bibinfo {year}
  {2017}{\natexlab{a}})}\BibitemShut {NoStop}%
\bibitem [{\citenamefont {Bruhat}\ \emph {et~al.}(2018)\citenamefont {Bruhat},
  \citenamefont {Cubaynes}, \citenamefont {Viennot}, \citenamefont {Dartiailh},
  \citenamefont {Desjardins}, \citenamefont {Cottet},\ and\ \citenamefont
  {Kontos}}]{Bruhat2018}%
  \BibitemOpen
  \bibfield  {author} {\bibinfo {author} {\bibfnamefont {L.~E}\ \bibnamefont
  {Bruhat}}, \bibinfo {author} {\bibfnamefont {T.}~\bibnamefont {Cubaynes}},
  \bibinfo {author} {\bibfnamefont {J.~J.}\ \bibnamefont {Viennot}}, \bibinfo
  {author} {\bibfnamefont {M.~C.}\ \bibnamefont {Dartiailh}}, \bibinfo {author}
  {\bibfnamefont {M.~M.}\ \bibnamefont {Desjardins}}, \bibinfo {author}
  {\bibfnamefont {A.}~\bibnamefont {Cottet}}, \ and\ \bibinfo {author}
  {\bibfnamefont {T.}~\bibnamefont {Kontos}},\ }\bibfield  {title} {\enquote
  {\bibinfo {title} {{Circuit QED with a quantum-dot charge qubit dressed by
  Cooper pairs}},}\ }\href {\doibase 10.1103/PhysRevB.98.155313} {\bibfield
  {journal} {\bibinfo  {journal} {Phys. Rev. B}\ }\textbf {\bibinfo {volume}
  {98}},\ \bibinfo {pages} {155313} (\bibinfo {year} {2018})}\BibitemShut
  {NoStop}%
\bibitem [{\citenamefont {Scarlino}\ \emph {et~al.}(2022)\citenamefont
  {Scarlino}, \citenamefont {Ungerer}, \citenamefont {van Woerkom},
  \citenamefont {Mancini}, \citenamefont {Stano}, \citenamefont {M\"uller},
  \citenamefont {Landig}, \citenamefont {Koski}, \citenamefont {Reichl},
  \citenamefont {Wegscheider}, \citenamefont {Ihn}, \citenamefont {Ensslin},\
  and\ \citenamefont {Wallraff}}]{Scarlino2022}%
  \BibitemOpen
  \bibfield  {author} {\bibinfo {author} {\bibfnamefont {P.}~\bibnamefont
  {Scarlino}}, \bibinfo {author} {\bibfnamefont {J.~H.}\ \bibnamefont
  {Ungerer}}, \bibinfo {author} {\bibfnamefont {D.~J.}\ \bibnamefont {van
  Woerkom}}, \bibinfo {author} {\bibfnamefont {M.}~\bibnamefont {Mancini}},
  \bibinfo {author} {\bibfnamefont {P.}~\bibnamefont {Stano}}, \bibinfo
  {author} {\bibfnamefont {C.}~\bibnamefont {M\"uller}}, \bibinfo {author}
  {\bibfnamefont {A.~J.}\ \bibnamefont {Landig}}, \bibinfo {author}
  {\bibfnamefont {J.~V.}\ \bibnamefont {Koski}}, \bibinfo {author}
  {\bibfnamefont {C.}~\bibnamefont {Reichl}}, \bibinfo {author} {\bibfnamefont
  {W.}~\bibnamefont {Wegscheider}}, \bibinfo {author} {\bibfnamefont
  {T.}~\bibnamefont {Ihn}}, \bibinfo {author} {\bibfnamefont {K.}~\bibnamefont
  {Ensslin}}, \ and\ \bibinfo {author} {\bibfnamefont {A.}~\bibnamefont
  {Wallraff}},\ }\bibfield  {title} {\enquote {\bibinfo {title} {{In situ
  Tuning of the Electric-Dipole Strength of a Double-Dot Charge Qubit:
  Charge-Noise Protection and Ultrastrong Coupling}},}\ }\href {\doibase
  10.1103/PhysRevX.12.031004} {\bibfield  {journal} {\bibinfo  {journal} {Phys.
  Rev. X}\ }\textbf {\bibinfo {volume} {12}},\ \bibinfo {pages} {031004}
  (\bibinfo {year} {2022})}\BibitemShut {NoStop}%
\bibitem [{\citenamefont {Yu}\ \emph {et~al.}(2023)\citenamefont {Yu},
  \citenamefont {Zihlmann}, \citenamefont {Abadillo-Uriel}, \citenamefont
  {Michal}, \citenamefont {Rambal}, \citenamefont {Niebojewski}, \citenamefont
  {Bedecarrats}, \citenamefont {Vinet}, \citenamefont {Dumur}, \citenamefont
  {Filippone}, \citenamefont {Bertrand}, \citenamefont {De~Franceschi},
  \citenamefont {Niquet},\ and\ \citenamefont {Maurand}}]{Yu2023}%
  \BibitemOpen
  \bibfield  {author} {\bibinfo {author} {\bibfnamefont {C.~X.}\ \bibnamefont
  {Yu}}, \bibinfo {author} {\bibfnamefont {S.}~\bibnamefont {Zihlmann}},
  \bibinfo {author} {\bibfnamefont {J.~C.}\ \bibnamefont {Abadillo-Uriel}},
  \bibinfo {author} {\bibfnamefont {V.~P.}\ \bibnamefont {Michal}}, \bibinfo
  {author} {\bibfnamefont {N.}~\bibnamefont {Rambal}}, \bibinfo {author}
  {\bibfnamefont {H.}~\bibnamefont {Niebojewski}}, \bibinfo {author}
  {\bibfnamefont {T.}~\bibnamefont {Bedecarrats}}, \bibinfo {author}
  {\bibfnamefont {M.}~\bibnamefont {Vinet}}, \bibinfo {author} {\bibfnamefont
  {{\'E}.}~\bibnamefont {Dumur}}, \bibinfo {author} {\bibfnamefont
  {M.}~\bibnamefont {Filippone}}, \bibinfo {author} {\bibfnamefont
  {B.}~\bibnamefont {Bertrand}}, \bibinfo {author} {\bibfnamefont
  {S.}~\bibnamefont {De~Franceschi}}, \bibinfo {author} {\bibfnamefont {Y.-M.}\
  \bibnamefont {Niquet}}, \ and\ \bibinfo {author} {\bibfnamefont
  {R.}~\bibnamefont {Maurand}},\ }\bibfield  {title} {\enquote {\bibinfo
  {title} {{Strong coupling between a photon and a hole spin in silicon}},}\
  }\href {\doibase 10.1038/s41565-023-01332-3} {\bibfield  {journal} {\bibinfo
  {journal} {Nature Nanotechnology}\ }\textbf {\bibinfo {volume} {18}},\
  \bibinfo {pages} {741--746} (\bibinfo {year} {2023})}\BibitemShut {NoStop}%
\bibitem [{\citenamefont {Nilsson}\ \emph {et~al.}(2016)\citenamefont
  {Nilsson}, \citenamefont {Namazi}, \citenamefont {Lehmann}, \citenamefont
  {Leijnse}, \citenamefont {Dick},\ and\ \citenamefont
  {Thelander}}]{Nilsson2016}%
  \BibitemOpen
  \bibfield  {author} {\bibinfo {author} {\bibfnamefont {M.}~\bibnamefont
  {Nilsson}}, \bibinfo {author} {\bibfnamefont {L.}~\bibnamefont {Namazi}},
  \bibinfo {author} {\bibfnamefont {S.}~\bibnamefont {Lehmann}}, \bibinfo
  {author} {\bibfnamefont {M.}~\bibnamefont {Leijnse}}, \bibinfo {author}
  {\bibfnamefont {K.~A.}\ \bibnamefont {Dick}}, \ and\ \bibinfo {author}
  {\bibfnamefont {C.}~\bibnamefont {Thelander}},\ }\bibfield  {title} {\enquote
  {\bibinfo {title} {{Single-electron transport in InAs nanowire quantum dots
  formed by crystal phase engineering}},}\ }\href {\doibase
  10.1103/PhysRevB.93.195422} {\bibfield  {journal} {\bibinfo  {journal} {Phys.
  Rev. B}\ }\textbf {\bibinfo {volume} {93}},\ \bibinfo {pages} {195422}
  (\bibinfo {year} {2016})}\BibitemShut {NoStop}%
\bibitem [{\citenamefont {Barker}\ \emph {et~al.}(2019)\citenamefont {Barker},
  \citenamefont {Lehmann}, \citenamefont {Namazi}, \citenamefont {Nilsson},
  \citenamefont {Thelander}, \citenamefont {Dick},\ and\ \citenamefont
  {Maisi}}]{Barker2019}%
  \BibitemOpen
  \bibfield  {author} {\bibinfo {author} {\bibfnamefont {D.}~\bibnamefont
  {Barker}}, \bibinfo {author} {\bibfnamefont {S.}~\bibnamefont {Lehmann}},
  \bibinfo {author} {\bibfnamefont {L.}~\bibnamefont {Namazi}}, \bibinfo
  {author} {\bibfnamefont {M.}~\bibnamefont {Nilsson}}, \bibinfo {author}
  {\bibfnamefont {C.}~\bibnamefont {Thelander}}, \bibinfo {author}
  {\bibfnamefont {K.~A.}\ \bibnamefont {Dick}}, \ and\ \bibinfo {author}
  {\bibfnamefont {V.~F.}\ \bibnamefont {Maisi}},\ }\bibfield  {title} {\enquote
  {\bibinfo {title} {{{Individually addressable double quantum dots formed with
  nanowire polytypes and identified by epitaxial markers}}},}\ }\href {\doibase
  10.1063/1.5089275} {\bibfield  {journal} {\bibinfo  {journal} {Applied
  Physics Letters}\ }\textbf {\bibinfo {volume} {114}},\ \bibinfo {pages}
  {183502} (\bibinfo {year} {2019})}\BibitemShut {NoStop}%
\bibitem [{\citenamefont {Masluk}\ \emph {et~al.}(2012)\citenamefont {Masluk},
  \citenamefont {Pop}, \citenamefont {Kamal}, \citenamefont {Minev},\ and\
  \citenamefont {Devoret}}]{Masluk2012}%
  \BibitemOpen
  \bibfield  {author} {\bibinfo {author} {\bibfnamefont {N.~A.}\ \bibnamefont
  {Masluk}}, \bibinfo {author} {\bibfnamefont {I.~M.}\ \bibnamefont {Pop}},
  \bibinfo {author} {\bibfnamefont {A.}~\bibnamefont {Kamal}}, \bibinfo
  {author} {\bibfnamefont {Z.~K.}\ \bibnamefont {Minev}}, \ and\ \bibinfo
  {author} {\bibfnamefont {M.~H.}\ \bibnamefont {Devoret}},\ }\bibfield
  {title} {\enquote {\bibinfo {title} {{Microwave Characterization of Josephson
  Junction Arrays: Implementing a Low Loss Superinductance}},}\ }\href
  {\doibase 10.1103/PhysRevLett.109.137002} {\bibfield  {journal} {\bibinfo
  {journal} {Phys. Rev. Lett.}\ }\textbf {\bibinfo {volume} {109}},\ \bibinfo
  {pages} {137002} (\bibinfo {year} {2012})}\BibitemShut {NoStop}%
\bibitem [{\citenamefont {Ranni}\ \emph {et~al.}(2023)\citenamefont {Ranni},
  \citenamefont {Havir}, \citenamefont {Haldar},\ and\ \citenamefont
  {Maisi}}]{Ranni2023}%
  \BibitemOpen
  \bibfield  {author} {\bibinfo {author} {\bibfnamefont {A.}~\bibnamefont
  {Ranni}}, \bibinfo {author} {\bibfnamefont {H.}~\bibnamefont {Havir}},
  \bibinfo {author} {\bibfnamefont {S.}~\bibnamefont {Haldar}}, \ and\ \bibinfo
  {author} {\bibfnamefont {V.~F.}\ \bibnamefont {Maisi}},\ }\href@noop {}
  {\enquote {\bibinfo {title} {{High Impedance Josephson Junction Resonators in
  the Transmission Line Geometry}},}\ } (\bibinfo {year} {2023}),\ \Eprint
  {http://arxiv.org/abs/2306.12701} {arXiv:2306.12701 [cond-mat.mes-hall]}
  \BibitemShut {NoStop}%
\bibitem [{\citenamefont {J{\"u}nger}\ \emph {et~al.}(2019)\citenamefont
  {J{\"u}nger}, \citenamefont {Baumgartner}, \citenamefont {Delagrange},
  \citenamefont {Chevallier}, \citenamefont {Lehmann}, \citenamefont {Nilsson},
  \citenamefont {Dick}, \citenamefont {Thelander},\ and\ \citenamefont
  {Sch{\"o}nenberger}}]{Junger2019}%
  \BibitemOpen
  \bibfield  {author} {\bibinfo {author} {\bibfnamefont {C.}~\bibnamefont
  {J{\"u}nger}}, \bibinfo {author} {\bibfnamefont {A.}~\bibnamefont
  {Baumgartner}}, \bibinfo {author} {\bibfnamefont {R.}~\bibnamefont
  {Delagrange}}, \bibinfo {author} {\bibfnamefont {D.}~\bibnamefont
  {Chevallier}}, \bibinfo {author} {\bibfnamefont {S.}~\bibnamefont {Lehmann}},
  \bibinfo {author} {\bibfnamefont {M.}~\bibnamefont {Nilsson}}, \bibinfo
  {author} {\bibfnamefont {K.~A.}\ \bibnamefont {Dick}}, \bibinfo {author}
  {\bibfnamefont {C.}~\bibnamefont {Thelander}}, \ and\ \bibinfo {author}
  {\bibfnamefont {C.}~\bibnamefont {Sch{\"o}nenberger}},\ }\bibfield  {title}
  {\enquote {\bibinfo {title} {{Spectroscopy of the superconducting proximity
  effect in nanowires using integrated quantum dots}},}\ }\href {\doibase
  10.1038/s42005-019-0162-4} {\bibfield  {journal} {\bibinfo  {journal}
  {Communications Physics}\ }\textbf {\bibinfo {volume} {2}},\ \bibinfo {pages}
  {76} (\bibinfo {year} {2019})}\BibitemShut {NoStop}%
\bibitem [{\citenamefont {Jünger}\ \emph {et~al.}(2021)\citenamefont
  {Jünger}, \citenamefont {Lehmann}, \citenamefont {Dick}, \citenamefont
  {Thelander}, \citenamefont {Schönenberger},\ and\ \citenamefont
  {Baumgartner}}]{Junger2021}%
  \BibitemOpen
  \bibfield  {author} {\bibinfo {author} {\bibfnamefont {C.}~\bibnamefont
  {Jünger}}, \bibinfo {author} {\bibfnamefont {S.}~\bibnamefont {Lehmann}},
  \bibinfo {author} {\bibfnamefont {K.~A.}\ \bibnamefont {Dick}}, \bibinfo
  {author} {\bibfnamefont {C.}~\bibnamefont {Thelander}}, \bibinfo {author}
  {\bibfnamefont {C.}~\bibnamefont {Schönenberger}}, \ and\ \bibinfo {author}
  {\bibfnamefont {A.}~\bibnamefont {Baumgartner}},\ }\href@noop {} {\enquote
  {\bibinfo {title} {{Intermediate states in Andreev bound state fusion}},}\ }
  (\bibinfo {year} {2021}),\ \Eprint {http://arxiv.org/abs/2111.00651}
  {arXiv:2111.00651 [cond-mat.mes-hall]} \BibitemShut {NoStop}%
\bibitem [{\citenamefont {Ungerer}\ \emph {et~al.}(2023)\citenamefont
  {Ungerer}, \citenamefont {Pally}, \citenamefont {Kononov}, \citenamefont
  {Lehmann}, \citenamefont {Ridderbos}, \citenamefont {Thelander},
  \citenamefont {Dick}, \citenamefont {Maisi}, \citenamefont {Scarlino},
  \citenamefont {Baumgartner},\ and\ \citenamefont
  {Schönenberger}}]{Ungerer2023}%
  \BibitemOpen
  \bibfield  {author} {\bibinfo {author} {\bibfnamefont {J.~H.}\ \bibnamefont
  {Ungerer}}, \bibinfo {author} {\bibfnamefont {A.}~\bibnamefont {Pally}},
  \bibinfo {author} {\bibfnamefont {A.}~\bibnamefont {Kononov}}, \bibinfo
  {author} {\bibfnamefont {S.}~\bibnamefont {Lehmann}}, \bibinfo {author}
  {\bibfnamefont {J.}~\bibnamefont {Ridderbos}}, \bibinfo {author}
  {\bibfnamefont {C.}~\bibnamefont {Thelander}}, \bibinfo {author}
  {\bibfnamefont {K.~A.}\ \bibnamefont {Dick}}, \bibinfo {author}
  {\bibfnamefont {V.~F.}\ \bibnamefont {Maisi}}, \bibinfo {author}
  {\bibfnamefont {P.}~\bibnamefont {Scarlino}}, \bibinfo {author}
  {\bibfnamefont {A.}~\bibnamefont {Baumgartner}}, \ and\ \bibinfo {author}
  {\bibfnamefont {C.}~\bibnamefont {Schönenberger}},\ }\href@noop {} {\enquote
  {\bibinfo {title} {{Strong coupling between a microwave photon and a
  singlet-triplet qubit}},}\ } (\bibinfo {year} {2023}),\ \Eprint
  {http://arxiv.org/abs/2303.16825} {arXiv:2303.16825 [cond-mat.mes-hall]}
  \BibitemShut {NoStop}%
\bibitem [{\citenamefont {Havir}\ \emph {et~al.}(2023)\citenamefont {Havir},
  \citenamefont {Haldar}, \citenamefont {Khan}, \citenamefont {Lehmann},
  \citenamefont {Dick}, \citenamefont {Thelander}, \citenamefont {Samuelsson},\
  and\ \citenamefont {Maisi}}]{Havir2023}%
  \BibitemOpen
  \bibfield  {author} {\bibinfo {author} {\bibfnamefont {H.}~\bibnamefont
  {Havir}}, \bibinfo {author} {\bibfnamefont {S.}~\bibnamefont {Haldar}},
  \bibinfo {author} {\bibfnamefont {W.}~\bibnamefont {Khan}}, \bibinfo {author}
  {\bibfnamefont {S.}~\bibnamefont {Lehmann}}, \bibinfo {author} {\bibfnamefont
  {K.~A.}\ \bibnamefont {Dick}}, \bibinfo {author} {\bibfnamefont
  {C.}~\bibnamefont {Thelander}}, \bibinfo {author} {\bibfnamefont
  {P.}~\bibnamefont {Samuelsson}}, \ and\ \bibinfo {author} {\bibfnamefont
  {V.~F.}\ \bibnamefont {Maisi}},\ }\href@noop {} {\enquote {\bibinfo {title}
  {{Quantum Dot Source-Drain Transport Response at Microwave Frequencies}},}\ }
  (\bibinfo {year} {2023}),\ \Eprint {http://arxiv.org/abs/2303.13048}
  {arXiv:2303.13048 [cond-mat.mes-hall]} \BibitemShut {NoStop}%
\bibitem [{\citenamefont {Göppl}\ \emph {et~al.}(2008)\citenamefont {Göppl},
  \citenamefont {Fragner}, \citenamefont {Baur}, \citenamefont {Bianchetti},
  \citenamefont {Filipp}, \citenamefont {Fink}, \citenamefont {Leek},
  \citenamefont {Puebla}, \citenamefont {Steffen},\ and\ \citenamefont
  {Wallraff}}]{Goppl2008}%
  \BibitemOpen
  \bibfield  {author} {\bibinfo {author} {\bibfnamefont {M.}~\bibnamefont
  {Göppl}}, \bibinfo {author} {\bibfnamefont {A.}~\bibnamefont {Fragner}},
  \bibinfo {author} {\bibfnamefont {M.}~\bibnamefont {Baur}}, \bibinfo {author}
  {\bibfnamefont {R.}~\bibnamefont {Bianchetti}}, \bibinfo {author}
  {\bibfnamefont {S.}~\bibnamefont {Filipp}}, \bibinfo {author} {\bibfnamefont
  {J.~M.}\ \bibnamefont {Fink}}, \bibinfo {author} {\bibfnamefont {P.~J.}\
  \bibnamefont {Leek}}, \bibinfo {author} {\bibfnamefont {G.}~\bibnamefont
  {Puebla}}, \bibinfo {author} {\bibfnamefont {L.}~\bibnamefont {Steffen}}, \
  and\ \bibinfo {author} {\bibfnamefont {A.}~\bibnamefont {Wallraff}},\
  }\bibfield  {title} {\enquote {\bibinfo {title} {{{Coplanar waveguide
  resonators for circuit quantum electrodynamics}}},}\ }\href {\doibase
  10.1063/1.3010859} {\bibfield  {journal} {\bibinfo  {journal} {Journal of
  Applied Physics}\ }\textbf {\bibinfo {volume} {104}},\ \bibinfo {pages}
  {113904} (\bibinfo {year} {2008})}\BibitemShut {NoStop}%
\bibitem [{\citenamefont {Haldar}\ \emph {et~al.}(2023)\citenamefont {Haldar},
  \citenamefont {Havir}, \citenamefont {Khan}, \citenamefont {Lehmann},
  \citenamefont {Thelander}, \citenamefont {Dick},\ and\ \citenamefont
  {Maisi}}]{Haldar2023}%
  \BibitemOpen
  \bibfield  {author} {\bibinfo {author} {\bibfnamefont {S.}~\bibnamefont
  {Haldar}}, \bibinfo {author} {\bibfnamefont {H.}~\bibnamefont {Havir}},
  \bibinfo {author} {\bibfnamefont {W.}~\bibnamefont {Khan}}, \bibinfo {author}
  {\bibfnamefont {S.}~\bibnamefont {Lehmann}}, \bibinfo {author} {\bibfnamefont
  {C.}~\bibnamefont {Thelander}}, \bibinfo {author} {\bibfnamefont {K.~A.}\
  \bibnamefont {Dick}}, \ and\ \bibinfo {author} {\bibfnamefont {V.~F.}\
  \bibnamefont {Maisi}},\ }\bibfield  {title} {\enquote {\bibinfo {title}
  {{Energetics of Microwaves Probed by Double Quantum Dot Absorption}},}\
  }\href {\doibase 10.1103/PhysRevLett.130.087003} {\bibfield  {journal}
  {\bibinfo  {journal} {Phys. Rev. Lett.}\ }\textbf {\bibinfo {volume} {130}},\
  \bibinfo {pages} {087003} (\bibinfo {year} {2023})}\BibitemShut {NoStop}%
\bibitem [{\citenamefont {Khan}\ \emph {et~al.}(2021)\citenamefont {Khan},
  \citenamefont {Potts}, \citenamefont {Lehmann}, \citenamefont {Thelander},
  \citenamefont {Dick}, \citenamefont {Samuelsson},\ and\ \citenamefont
  {Maisi}}]{Khan2021}%
  \BibitemOpen
  \bibfield  {author} {\bibinfo {author} {\bibfnamefont {W.}~\bibnamefont
  {Khan}}, \bibinfo {author} {\bibfnamefont {P.~P.}\ \bibnamefont {Potts}},
  \bibinfo {author} {\bibfnamefont {S.}~\bibnamefont {Lehmann}}, \bibinfo
  {author} {\bibfnamefont {C.}~\bibnamefont {Thelander}}, \bibinfo {author}
  {\bibfnamefont {K.~A.}\ \bibnamefont {Dick}}, \bibinfo {author}
  {\bibfnamefont {P.}~\bibnamefont {Samuelsson}}, \ and\ \bibinfo {author}
  {\bibfnamefont {V.~F.}\ \bibnamefont {Maisi}},\ }\bibfield  {title} {\enquote
  {\bibinfo {title} {{Efficient and continuous microwave photoconversion in
  hybrid cavity-semiconductor nanowire double quantum dot diodes}},}\ }\href
  {\doibase 10.1038/s41467-021-25446-1} {\bibfield  {journal} {\bibinfo
  {journal} {Nature Communications}\ }\textbf {\bibinfo {volume} {12}},\
  \bibinfo {pages} {5130} (\bibinfo {year} {2021})}\BibitemShut {NoStop}%
\bibitem [{\citenamefont {Frey}\ \emph {et~al.}(2012)\citenamefont {Frey},
  \citenamefont {Leek}, \citenamefont {Beck}, \citenamefont {Blais},
  \citenamefont {Ihn}, \citenamefont {Ensslin},\ and\ \citenamefont
  {Wallraff}}]{Frey2012}%
  \BibitemOpen
  \bibfield  {author} {\bibinfo {author} {\bibfnamefont {T.}~\bibnamefont
  {Frey}}, \bibinfo {author} {\bibfnamefont {P.~J.}\ \bibnamefont {Leek}},
  \bibinfo {author} {\bibfnamefont {M.}~\bibnamefont {Beck}}, \bibinfo {author}
  {\bibfnamefont {A.}~\bibnamefont {Blais}}, \bibinfo {author} {\bibfnamefont
  {T.}~\bibnamefont {Ihn}}, \bibinfo {author} {\bibfnamefont {K.}~\bibnamefont
  {Ensslin}}, \ and\ \bibinfo {author} {\bibfnamefont {A.}~\bibnamefont
  {Wallraff}},\ }\bibfield  {title} {\enquote {\bibinfo {title} {{Dipole
  Coupling of a Double Quantum Dot to a Microwave Resonator}},}\ }\href
  {\doibase 10.1103/PhysRevLett.108.046807} {\bibfield  {journal} {\bibinfo
  {journal} {Phys. Rev. Lett.}\ }\textbf {\bibinfo {volume} {108}},\ \bibinfo
  {pages} {046807} (\bibinfo {year} {2012})}\BibitemShut {NoStop}%
\bibitem [{\citenamefont {Petta}\ \emph {et~al.}(2005)\citenamefont {Petta},
  \citenamefont {Johnson}, \citenamefont {Taylor}, \citenamefont {Laird},
  \citenamefont {Yacoby}, \citenamefont {Lukin}, \citenamefont {Marcus},
  \citenamefont {Hanson},\ and\ \citenamefont {Gossard}}]{Petta2005}%
  \BibitemOpen
  \bibfield  {author} {\bibinfo {author} {\bibfnamefont {J.~R.}\ \bibnamefont
  {Petta}}, \bibinfo {author} {\bibfnamefont {A.~C.}\ \bibnamefont {Johnson}},
  \bibinfo {author} {\bibfnamefont {J.~M.}\ \bibnamefont {Taylor}}, \bibinfo
  {author} {\bibfnamefont {E.~A.}\ \bibnamefont {Laird}}, \bibinfo {author}
  {\bibfnamefont {A.}~\bibnamefont {Yacoby}}, \bibinfo {author} {\bibfnamefont
  {M.~D.}\ \bibnamefont {Lukin}}, \bibinfo {author} {\bibfnamefont {C.~M.}\
  \bibnamefont {Marcus}}, \bibinfo {author} {\bibfnamefont {M.~P.}\
  \bibnamefont {Hanson}}, \ and\ \bibinfo {author} {\bibfnamefont {A.~C.}\
  \bibnamefont {Gossard}},\ }\bibfield  {title} {\enquote {\bibinfo {title}
  {{Coherent Manipulation of Coupled Electron Spins in Semiconductor Quantum
  Dots}},}\ }\href {\doibase 10.1126/science.1116955} {\bibfield  {journal}
  {\bibinfo  {journal} {Science}\ }\textbf {\bibinfo {volume} {309}},\ \bibinfo
  {pages} {2180--2184} (\bibinfo {year} {2005})}\BibitemShut {NoStop}%
\bibitem [{\citenamefont {Childress}\ \emph {et~al.}(2004)\citenamefont
  {Childress}, \citenamefont {S\o{}rensen},\ and\ \citenamefont
  {Lukin}}]{Childress2004}%
  \BibitemOpen
  \bibfield  {author} {\bibinfo {author} {\bibfnamefont {L.}~\bibnamefont
  {Childress}}, \bibinfo {author} {\bibfnamefont {A.~S.}\ \bibnamefont
  {S\o{}rensen}}, \ and\ \bibinfo {author} {\bibfnamefont {M.~D.}\ \bibnamefont
  {Lukin}},\ }\bibfield  {title} {\enquote {\bibinfo {title} {{Mesoscopic
  cavity quantum electrodynamics with quantum dots}},}\ }\href {\doibase
  10.1103/PhysRevA.69.042302} {\bibfield  {journal} {\bibinfo  {journal} {Phys.
  Rev. A}\ }\textbf {\bibinfo {volume} {69}},\ \bibinfo {pages} {042302}
  (\bibinfo {year} {2004})}\BibitemShut {NoStop}%
\bibitem [{\citenamefont {Blais}\ \emph {et~al.}(2021)\citenamefont {Blais},
  \citenamefont {Grimsmo}, \citenamefont {Girvin},\ and\ \citenamefont
  {Wallraff}}]{Blais2021}%
  \BibitemOpen
  \bibfield  {author} {\bibinfo {author} {\bibfnamefont {A.}~\bibnamefont
  {Blais}}, \bibinfo {author} {\bibfnamefont {A.~L.}\ \bibnamefont {Grimsmo}},
  \bibinfo {author} {\bibfnamefont {S.~M.}\ \bibnamefont {Girvin}}, \ and\
  \bibinfo {author} {\bibfnamefont {A.}~\bibnamefont {Wallraff}},\ }\bibfield
  {title} {\enquote {\bibinfo {title} {{Circuit quantum electrodynamics}},}\
  }\href {\doibase 10.1103/RevModPhys.93.025005} {\bibfield  {journal}
  {\bibinfo  {journal} {Rev. Mod. Phys.}\ }\textbf {\bibinfo {volume} {93}},\
  \bibinfo {pages} {025005} (\bibinfo {year} {2021})}\BibitemShut {NoStop}%
\bibitem [{\citenamefont {Dial}\ \emph {et~al.}(2013)\citenamefont {Dial},
  \citenamefont {Shulman}, \citenamefont {Harvey}, \citenamefont {Bluhm},
  \citenamefont {Umansky},\ and\ \citenamefont {Yacoby}}]{Dial2013}%
  \BibitemOpen
  \bibfield  {author} {\bibinfo {author} {\bibfnamefont {O.~E.}\ \bibnamefont
  {Dial}}, \bibinfo {author} {\bibfnamefont {M.~D.}\ \bibnamefont {Shulman}},
  \bibinfo {author} {\bibfnamefont {S.~P.}\ \bibnamefont {Harvey}}, \bibinfo
  {author} {\bibfnamefont {H.}~\bibnamefont {Bluhm}}, \bibinfo {author}
  {\bibfnamefont {V.}~\bibnamefont {Umansky}}, \ and\ \bibinfo {author}
  {\bibfnamefont {A.}~\bibnamefont {Yacoby}},\ }\bibfield  {title} {\enquote
  {\bibinfo {title} {{Charge Noise Spectroscopy Using Coherent Exchange
  Oscillations in a Singlet-Triplet Qubit}},}\ }\href {\doibase
  10.1103/PhysRevLett.110.146804} {\bibfield  {journal} {\bibinfo  {journal}
  {Phys. Rev. Lett.}\ }\textbf {\bibinfo {volume} {110}},\ \bibinfo {pages}
  {146804} (\bibinfo {year} {2013})}\BibitemShut {NoStop}%
\bibitem [{\citenamefont {Bermeister}\ \emph {et~al.}(2014)\citenamefont
  {Bermeister}, \citenamefont {Keith},\ and\ \citenamefont
  {Culcer}}]{Bermeister2014}%
  \BibitemOpen
  \bibfield  {author} {\bibinfo {author} {\bibfnamefont {A.}~\bibnamefont
  {Bermeister}}, \bibinfo {author} {\bibfnamefont {D.}~\bibnamefont {Keith}}, \
  and\ \bibinfo {author} {\bibfnamefont {D.}~\bibnamefont {Culcer}},\
  }\bibfield  {title} {\enquote {\bibinfo {title} {{{Charge noise, spin-orbit
  coupling, and dephasing of single-spin qubits}}},}\ }\href {\doibase
  10.1063/1.4901162} {\bibfield  {journal} {\bibinfo  {journal} {Applied
  Physics Letters}\ }\textbf {\bibinfo {volume} {105}},\ \bibinfo {pages}
  {192102} (\bibinfo {year} {2014})}\BibitemShut {NoStop}%
\bibitem [{\citenamefont {Petersson}\ \emph {et~al.}(2010)\citenamefont
  {Petersson}, \citenamefont {Petta}, \citenamefont {Lu},\ and\ \citenamefont
  {Gossard}}]{Petersson2010}%
  \BibitemOpen
  \bibfield  {author} {\bibinfo {author} {\bibfnamefont {K.~D.}\ \bibnamefont
  {Petersson}}, \bibinfo {author} {\bibfnamefont {J.~R.}\ \bibnamefont
  {Petta}}, \bibinfo {author} {\bibfnamefont {H.}~\bibnamefont {Lu}}, \ and\
  \bibinfo {author} {\bibfnamefont {A.~C.}\ \bibnamefont {Gossard}},\
  }\bibfield  {title} {\enquote {\bibinfo {title} {{Quantum Coherence in a
  One-Electron Semiconductor Charge Qubit}},}\ }\href {\doibase
  10.1103/PhysRevLett.105.246804} {\bibfield  {journal} {\bibinfo  {journal}
  {Phys. Rev. Lett.}\ }\textbf {\bibinfo {volume} {105}},\ \bibinfo {pages}
  {246804} (\bibinfo {year} {2010})}\BibitemShut {NoStop}%
\bibitem [{\citenamefont {Vion}\ \emph {et~al.}(2002)\citenamefont {Vion},
  \citenamefont {Aassime}, \citenamefont {Cottet}, \citenamefont {Joyez},
  \citenamefont {Pothier}, \citenamefont {Urbina}, \citenamefont {Esteve},\
  and\ \citenamefont {Devoret}}]{Vion2002}%
  \BibitemOpen
  \bibfield  {author} {\bibinfo {author} {\bibfnamefont {D.}~\bibnamefont
  {Vion}}, \bibinfo {author} {\bibfnamefont {A.}~\bibnamefont {Aassime}},
  \bibinfo {author} {\bibfnamefont {A.}~\bibnamefont {Cottet}}, \bibinfo
  {author} {\bibfnamefont {P.}~\bibnamefont {Joyez}}, \bibinfo {author}
  {\bibfnamefont {H.}~\bibnamefont {Pothier}}, \bibinfo {author} {\bibfnamefont
  {C.}~\bibnamefont {Urbina}}, \bibinfo {author} {\bibfnamefont
  {D.}~\bibnamefont {Esteve}}, \ and\ \bibinfo {author} {\bibfnamefont {M.~H.}\
  \bibnamefont {Devoret}},\ }\bibfield  {title} {\enquote {\bibinfo {title}
  {{Manipulating the Quantum State of an Electrical Circuit}},}\ }\href
  {\doibase 10.1126/science.1069372} {\bibfield  {journal} {\bibinfo  {journal}
  {Science}\ }\textbf {\bibinfo {volume} {296}},\ \bibinfo {pages} {886--889}
  (\bibinfo {year} {2002})}\BibitemShut {NoStop}%
\bibitem [{\citenamefont {Nakamura}\ \emph {et~al.}(1999)\citenamefont
  {Nakamura}, \citenamefont {Pashkin},\ and\ \citenamefont
  {Tsai}}]{Nakamura1999}%
  \BibitemOpen
  \bibfield  {author} {\bibinfo {author} {\bibfnamefont {Y.}~\bibnamefont
  {Nakamura}}, \bibinfo {author} {\bibfnamefont {Yu.~A.}\ \bibnamefont
  {Pashkin}}, \ and\ \bibinfo {author} {\bibfnamefont {J.~S.}\ \bibnamefont
  {Tsai}},\ }\bibfield  {title} {\enquote {\bibinfo {title} {{Coherent control
  of macroscopic quantum states in a single-Cooper-pair box}},}\ }\href
  {\doibase 10.1038/19718} {\bibfield  {journal} {\bibinfo  {journal} {Nature}\
  }\textbf {\bibinfo {volume} {398}},\ \bibinfo {pages} {786--788} (\bibinfo
  {year} {1999})}\BibitemShut {NoStop}%
\bibitem [{\citenamefont {Medford}\ \emph {et~al.}(2013)\citenamefont
  {Medford}, \citenamefont {Beil}, \citenamefont {Taylor}, \citenamefont
  {Rashba}, \citenamefont {Lu}, \citenamefont {Gossard},\ and\ \citenamefont
  {Marcus}}]{Medford2013}%
  \BibitemOpen
  \bibfield  {author} {\bibinfo {author} {\bibfnamefont {J.}~\bibnamefont
  {Medford}}, \bibinfo {author} {\bibfnamefont {J.}~\bibnamefont {Beil}},
  \bibinfo {author} {\bibfnamefont {J.~M.}\ \bibnamefont {Taylor}}, \bibinfo
  {author} {\bibfnamefont {E.~I.}\ \bibnamefont {Rashba}}, \bibinfo {author}
  {\bibfnamefont {H.}~\bibnamefont {Lu}}, \bibinfo {author} {\bibfnamefont
  {A.~C.}\ \bibnamefont {Gossard}}, \ and\ \bibinfo {author} {\bibfnamefont
  {C.~M.}\ \bibnamefont {Marcus}},\ }\bibfield  {title} {\enquote {\bibinfo
  {title} {{Quantum-Dot-Based Resonant Exchange Qubit}},}\ }\href {\doibase
  10.1103/PhysRevLett.111.050501} {\bibfield  {journal} {\bibinfo  {journal}
  {Phys. Rev. Lett.}\ }\textbf {\bibinfo {volume} {111}},\ \bibinfo {pages}
  {050501} (\bibinfo {year} {2013})}\BibitemShut {NoStop}%
\bibitem [{\citenamefont {Thorgrimsson}\ \emph {et~al.}(2017)\citenamefont
  {Thorgrimsson}, \citenamefont {Kim}, \citenamefont {Yang}, \citenamefont
  {Smith}, \citenamefont {Simmons}, \citenamefont {Ward}, \citenamefont
  {Foote}, \citenamefont {Corrigan}, \citenamefont {Savage}, \citenamefont
  {Lagally}, \citenamefont {Friesen}, \citenamefont {Coppersmith},\ and\
  \citenamefont {Eriksson}}]{Thorgrimsson2017}%
  \BibitemOpen
  \bibfield  {author} {\bibinfo {author} {\bibfnamefont {B.}~\bibnamefont
  {Thorgrimsson}}, \bibinfo {author} {\bibfnamefont {D.}~\bibnamefont {Kim}},
  \bibinfo {author} {\bibfnamefont {Y.-C.}\ \bibnamefont {Yang}}, \bibinfo
  {author} {\bibfnamefont {L.~W.}\ \bibnamefont {Smith}}, \bibinfo {author}
  {\bibfnamefont {C.~B.}\ \bibnamefont {Simmons}}, \bibinfo {author}
  {\bibfnamefont {D.~R.}\ \bibnamefont {Ward}}, \bibinfo {author}
  {\bibfnamefont {R.~H.}\ \bibnamefont {Foote}}, \bibinfo {author}
  {\bibfnamefont {J.}~\bibnamefont {Corrigan}}, \bibinfo {author}
  {\bibfnamefont {D.~E.}\ \bibnamefont {Savage}}, \bibinfo {author}
  {\bibfnamefont {M.~G.}\ \bibnamefont {Lagally}}, \bibinfo {author}
  {\bibfnamefont {M.}~\bibnamefont {Friesen}}, \bibinfo {author} {\bibfnamefont
  {S.~N.}\ \bibnamefont {Coppersmith}}, \ and\ \bibinfo {author} {\bibfnamefont
  {M.~A.}\ \bibnamefont {Eriksson}},\ }\bibfield  {title} {\enquote {\bibinfo
  {title} {{Extending the coherence of a quantum dot hybrid qubit}},}\ }\href
  {\doibase 10.1038/s41534-017-0034-2} {\bibfield  {journal} {\bibinfo
  {journal} {npj Quantum Information}\ }\textbf {\bibinfo {volume} {3}},\
  \bibinfo {pages} {32} (\bibinfo {year} {2017})}\BibitemShut {NoStop}%
\bibitem [{\citenamefont {Fujisawa}\ \emph {et~al.}(1998)\citenamefont
  {Fujisawa}, \citenamefont {Oosterkamp}, \citenamefont {van~der Wiel},
  \citenamefont {Broer}, \citenamefont {Aguado}, \citenamefont {Tarucha},\ and\
  \citenamefont {Kouwenhoven}}]{fujisawa1998}%
  \BibitemOpen
  \bibfield  {author} {\bibinfo {author} {\bibfnamefont {T.}~\bibnamefont
  {Fujisawa}}, \bibinfo {author} {\bibfnamefont {T.~H.}\ \bibnamefont
  {Oosterkamp}}, \bibinfo {author} {\bibfnamefont {W.~G.}\ \bibnamefont
  {van~der Wiel}}, \bibinfo {author} {\bibfnamefont {B.~W.}\ \bibnamefont
  {Broer}}, \bibinfo {author} {\bibfnamefont {R.}~\bibnamefont {Aguado}},
  \bibinfo {author} {\bibfnamefont {S.}~\bibnamefont {Tarucha}}, \ and\
  \bibinfo {author} {\bibfnamefont {L.~P.}\ \bibnamefont {Kouwenhoven}},\
  }\bibfield  {title} {\enquote {\bibinfo {title} {{Spontaneous Emission
  Spectrum in Double Quantum Dot Devices}},}\ }\href {\doibase
  10.1126/science.282.5390.932} {\bibfield  {journal} {\bibinfo  {journal}
  {Science}\ }\textbf {\bibinfo {volume} {282}},\ \bibinfo {pages} {932--935}
  (\bibinfo {year} {1998})}\BibitemShut {NoStop}%
\bibitem [{\citenamefont {Hartke}\ \emph {et~al.}(2018)\citenamefont {Hartke},
  \citenamefont {Liu}, \citenamefont {Gullans},\ and\ \citenamefont
  {Petta}}]{hartke2018}%
  \BibitemOpen
  \bibfield  {author} {\bibinfo {author} {\bibfnamefont {T.~R.}\ \bibnamefont
  {Hartke}}, \bibinfo {author} {\bibfnamefont {Y.-Y.}\ \bibnamefont {Liu}},
  \bibinfo {author} {\bibfnamefont {M.~J.}\ \bibnamefont {Gullans}}, \ and\
  \bibinfo {author} {\bibfnamefont {J.~R.}\ \bibnamefont {Petta}},\ }\bibfield
  {title} {\enquote {\bibinfo {title} {{Microwave Detection of Electron-Phonon
  Interactions in a Cavity-Coupled Double Quantum Dot}},}\ }\href {\doibase
  10.1103/PhysRevLett.120.097701} {\bibfield  {journal} {\bibinfo  {journal}
  {Phys. Rev. Lett.}\ }\textbf {\bibinfo {volume} {120}},\ \bibinfo {pages}
  {097701} (\bibinfo {year} {2018})}\BibitemShut {NoStop}%
\bibitem [{\citenamefont {Hofmann}\ \emph {et~al.}(2020)\citenamefont
  {Hofmann}, \citenamefont {Karlewski}, \citenamefont {Heimes}, \citenamefont
  {Reichl}, \citenamefont {Wegscheider}, \citenamefont {Sch\"on}, \citenamefont
  {Ensslin}, \citenamefont {Ihn},\ and\ \citenamefont {Maisi}}]{Maisi2020}%
  \BibitemOpen
  \bibfield  {author} {\bibinfo {author} {\bibfnamefont {A.}~\bibnamefont
  {Hofmann}}, \bibinfo {author} {\bibfnamefont {C.}~\bibnamefont {Karlewski}},
  \bibinfo {author} {\bibfnamefont {A.}~\bibnamefont {Heimes}}, \bibinfo
  {author} {\bibfnamefont {C.}~\bibnamefont {Reichl}}, \bibinfo {author}
  {\bibfnamefont {W.}~\bibnamefont {Wegscheider}}, \bibinfo {author}
  {\bibfnamefont {G.}~\bibnamefont {Sch\"on}}, \bibinfo {author} {\bibfnamefont
  {K.}~\bibnamefont {Ensslin}}, \bibinfo {author} {\bibfnamefont
  {T.}~\bibnamefont {Ihn}}, \ and\ \bibinfo {author} {\bibfnamefont {V.~F.}\
  \bibnamefont {Maisi}},\ }\bibfield  {title} {\enquote {\bibinfo {title}
  {{Phonon spectral density in a GaAs/AlGaAs double quantum dot}},}\ }\href
  {\doibase 10.1103/PhysRevResearch.2.033230} {\bibfield  {journal} {\bibinfo
  {journal} {Phys. Rev. Res.}\ }\textbf {\bibinfo {volume} {2}},\ \bibinfo
  {pages} {033230} (\bibinfo {year} {2020})}\BibitemShut {NoStop}%
\bibitem [{\citenamefont {Mi}\ \emph {et~al.}(2017{\natexlab{b}})\citenamefont
  {Mi}, \citenamefont {Cady}, \citenamefont {Zajac}, \citenamefont {Stehlik},
  \citenamefont {Edge},\ and\ \citenamefont {Petta}}]{Mi2017b}%
  \BibitemOpen
  \bibfield  {author} {\bibinfo {author} {\bibfnamefont {X.}~\bibnamefont
  {Mi}}, \bibinfo {author} {\bibfnamefont {J.~V.}\ \bibnamefont {Cady}},
  \bibinfo {author} {\bibfnamefont {D.~M.}\ \bibnamefont {Zajac}}, \bibinfo
  {author} {\bibfnamefont {J.}~\bibnamefont {Stehlik}}, \bibinfo {author}
  {\bibfnamefont {L.~F.}\ \bibnamefont {Edge}}, \ and\ \bibinfo {author}
  {\bibfnamefont {J.~R.}\ \bibnamefont {Petta}},\ }\bibfield  {title} {\enquote
  {\bibinfo {title} {{{Circuit quantum electrodynamics architecture for
  gate-defined quantum dots in silicon}}},}\ }\href {\doibase
  10.1063/1.4974536} {\bibfield  {journal} {\bibinfo  {journal} {Applied
  Physics Letters}\ }\textbf {\bibinfo {volume} {110}},\ \bibinfo {pages}
  {043502} (\bibinfo {year} {2017}{\natexlab{b}})}\BibitemShut {NoStop}%
\bibitem [{\citenamefont {Guthrie}\ \emph {et~al.}(2022)\citenamefont
  {Guthrie}, \citenamefont {Satrya}, \citenamefont {Chang}, \citenamefont
  {Menczel}, \citenamefont {Nori},\ and\ \citenamefont {Pekola}}]{guthrie2022}%
  \BibitemOpen
  \bibfield  {author} {\bibinfo {author} {\bibfnamefont {A.}~\bibnamefont
  {Guthrie}}, \bibinfo {author} {\bibfnamefont {C.~D.}\ \bibnamefont {Satrya}},
  \bibinfo {author} {\bibfnamefont {Y.-C.}\ \bibnamefont {Chang}}, \bibinfo
  {author} {\bibfnamefont {P.}~\bibnamefont {Menczel}}, \bibinfo {author}
  {\bibfnamefont {F.}~\bibnamefont {Nori}}, \ and\ \bibinfo {author}
  {\bibfnamefont {J.~P.}\ \bibnamefont {Pekola}},\ }\bibfield  {title}
  {\enquote {\bibinfo {title} {{Cooper-Pair Box Coupled to Two Resonators: An
  Architecture for a Quantum Refrigerator}},}\ }\href {\doibase
  10.1103/PhysRevApplied.17.064022} {\bibfield  {journal} {\bibinfo  {journal}
  {Phys. Rev. Appl.}\ }\textbf {\bibinfo {volume} {17}},\ \bibinfo {pages}
  {064022} (\bibinfo {year} {2022})}\BibitemShut {NoStop}%
\bibitem [{\citenamefont {Golovach}\ \emph {et~al.}(2008)\citenamefont
  {Golovach}, \citenamefont {Khaetskii},\ and\ \citenamefont
  {Loss}}]{Golovach2008}%
  \BibitemOpen
  \bibfield  {author} {\bibinfo {author} {\bibfnamefont {V.~N.}\ \bibnamefont
  {Golovach}}, \bibinfo {author} {\bibfnamefont {A.}~\bibnamefont {Khaetskii}},
  \ and\ \bibinfo {author} {\bibfnamefont {D.}~\bibnamefont {Loss}},\
  }\bibfield  {title} {\enquote {\bibinfo {title} {{Spin relaxation at the
  singlet-triplet crossing in a quantum dot}},}\ }\href {\doibase
  10.1103/PhysRevB.77.045328} {\bibfield  {journal} {\bibinfo  {journal} {Phys.
  Rev. B}\ }\textbf {\bibinfo {volume} {77}},\ \bibinfo {pages} {045328}
  (\bibinfo {year} {2008})}\BibitemShut {NoStop}%
\bibitem [{\citenamefont {Roulleau}\ \emph {et~al.}(2011)\citenamefont
  {Roulleau}, \citenamefont {Baer}, \citenamefont {Choi}, \citenamefont
  {Molitor}, \citenamefont {G{\"u}ttinger}, \citenamefont {M{\"u}ller},
  \citenamefont {Dr{\"o}scher}, \citenamefont {Ensslin},\ and\ \citenamefont
  {Ihn}}]{Roulleau2011}%
  \BibitemOpen
  \bibfield  {author} {\bibinfo {author} {\bibfnamefont {P.}~\bibnamefont
  {Roulleau}}, \bibinfo {author} {\bibfnamefont {S.}~\bibnamefont {Baer}},
  \bibinfo {author} {\bibfnamefont {T.}~\bibnamefont {Choi}}, \bibinfo {author}
  {\bibfnamefont {F.}~\bibnamefont {Molitor}}, \bibinfo {author} {\bibfnamefont
  {J.}~\bibnamefont {G{\"u}ttinger}}, \bibinfo {author} {\bibfnamefont
  {T.}~\bibnamefont {M{\"u}ller}}, \bibinfo {author} {\bibfnamefont
  {S.}~\bibnamefont {Dr{\"o}scher}}, \bibinfo {author} {\bibfnamefont
  {K.}~\bibnamefont {Ensslin}}, \ and\ \bibinfo {author} {\bibfnamefont
  {T.}~\bibnamefont {Ihn}},\ }\bibfield  {title} {\enquote {\bibinfo {title}
  {{Coherent electron--phonon coupling in tailored quantum systems}},}\ }\href
  {\doibase 10.1038/ncomms1241} {\bibfield  {journal} {\bibinfo  {journal}
  {Nature Communications}\ }\textbf {\bibinfo {volume} {2}},\ \bibinfo {pages}
  {239} (\bibinfo {year} {2011})}\BibitemShut {NoStop}%
\bibitem [{\citenamefont {Gardiner}\ and\ \citenamefont
  {Zoller}(2004)}]{gardiner_book}%
  \BibitemOpen
  \bibfield  {author} {\bibinfo {author} {\bibfnamefont {C.~W.}\ \bibnamefont
  {Gardiner}}\ and\ \bibinfo {author} {\bibfnamefont {P.}~\bibnamefont
  {Zoller}},\ }\href@noop {} {\emph {\bibinfo {title} {{Quantum Noise: A
  Handbook of Markovian and Non-Markovian Quantum Stochastic Methods with
  Applications to Quantum Optics}}}},\ \bibinfo {edition} {3rd}\ ed.\ (\bibinfo
   {publisher} {Springer Berlin, Heidelberg},\ \bibinfo {year}
  {2004})\BibitemShut {NoStop}%
\end{thebibliography}%

\end{document}